\documentclass[sensors,article,accept,moreauthors,LaTeX,dvi2pdf]{mdpi}

\firstpage{1} 
\pubvolume{xx}
\issuenum{1}
\articlenumber{5}
\pubyear{2023}
\copyrightyear{2023}
\history{Received: date; Accepted: date; Published: date}

\usepackage{epsfig}

\Title{Li$_2$$^{100\textnormal{depl}}$MoO$_4$ Scintillating Bolometers for Rare-Event Search Experiments}

\Author{
I.C.~Bandac $^{1}$,
A.S.~Barabash $^{2}$,
L.~Berg\'e $^{3}$,
Yu.A.~Borovlev $^{4}$,
J.M.~Calvo-Mozota $^{1,5}$,
P.~Carniti $^{6}$,
M.~Chapellier $^{3}$,
I.~Dafinei $^{7}$,
F.A.~Danevich $^{8,9}$,
L.~Dumoulin $^{3}$,
F.~Ferri $^{10}$,
A.~Giuliani $^{3}$,
C.~Gotti $^{6}$,
Ph.~Gras $^{10}$,
V.D.~Grigorieva $^{4}$,
A.~Ianni $^{11}$,
H.~Khalife $^{10}$,
V.V.~Kobychev $^{8}$,
S.I.~Konovalov $^{2}$,
P.~Loaiza $^{3}$,
M.~Madhukuttan $^{3}$,
E.P.~Makarov $^{4}$,
P.~de~Marcillac $^{3}$,
S.~Marnieros $^{3}$,
C.A.~Marrache-Kikuchi $^{3}$,
M.~Martinez $^{12}$,
C.~Nones $^{10}$,
E.~Olivieri $^{3}$,
A.~Ortiz de Sol\'orzano $^{12}$,
G.~Pessina $^{6}$,
D.V.~Poda $^{3}$\orcidA{}*,
Th.~Redon $^{3}$,
J.A.~Scarpaci $^{3}$,
V.N.~Shlegel $^{4}$,
V.I.~Tretyak $^{8,11}$,
V.I.~Umatov $^{2}$,
M.M.~Zarytskyy $^{8}$,
and A.~Zolotarova $^{10}$}

\AuthorNames{Denys Poda and CROSS collaboration}

\address{%
$^{1}$ \quad Laboratorio Subterr\'aneo de Canfranc, 22880 Canfranc-Estaci\'on, Spain \\
$^{2}$ \quad National Research Centre Kurchatov Institute, Kurchatov Complex of Theoretical and Experimental Physics, 117218 Moscow, Russia \\
$^{3}$ \quad Universit\'{e} Paris-Saclay, CNRS/IN2P3, IJCLab, 91405 Orsay, France \\
$^{4}$ \quad Nikolaev Institute of Inorganic Chemistry, 630090 Novosibirsk, Russia \\
$^{5}$ \quad Escuela Superior de Ingenier\'ia y Tecnolog\'ia, Universidad Internacional de La Rioja, 26006 Logro\~no, Spain \\
$^{6}$ \quad INFN, Sezione di Milano Bicocca, I-20126 Milano, Italy \\
$^{7}$ \quad INFN, Sezione di Roma, I-00185 Rome, Italy \\
$^{8}$ \quad Institute for Nuclear Research of NASU, 03028 Kyiv, Ukraine \\
$^{9}$ \quad INFN Sezione di Roma Tor Vergata, I-00133 Rome, Italy \\
$^{10}$ \quad IRFU, CEA, Universit\'e Paris-Saclay, F-91191 Gif-sur-Yvette, France \\
$^{11}$ \quad INFN, Laboratori Nazionali del Gran Sasso, I-67100 Assergi (AQ), Italy \\
$^{12}$ \quad Fundaci\'on ARAID \& Centro de Astropart\'iculas y F\'isica de Altas Energ\'ias, Universidad de Zaragoza, 50009 Zaragoza, Spain}

\corres{Correspondence: denys.poda@ijclab.in2p3.fr}

\abstract{We report on the development of scintillating bolometers based on lithium molybdate crystals containing molybdenum depleted in the double-$\beta$ active isotope $^{100}$Mo (Li$_2$$^{100\textnormal{depl}}$MoO$_4$). We used two Li$_2$$^{100\textnormal{depl}}$MoO$_4$ cubic samples, 45 mm side and 0.28 kg each, produced following purification and crystallization protocols developed for double-$\beta$ search experiments with $^{100}$Mo-enriched Li$_2$MoO$_4$ crystals. Bolometric Ge detectors were utilized to register scintillation photons emitted by the Li$_2$$^{100\textnormal{depl}}$MoO$_4$ crystal scintillators. The measurements were performed in the CROSS cryogenic set-up at the Canfranc underground laboratory (Spain). We observed that the Li$_2$$^{100\textnormal{depl}}$MoO$_4$ scintillating bolometers are characterized by excellent spectrometric performance ($\sim$3--6 keV FWHM at 0.24--2.6 MeV $\gamma$'s), moderate scintillation signal ($\sim$0.3--0.6 keV/MeV depending on light collection conditions) and high radiopurity ($^{228}$Th and $^{226}$Ra activities are below a few $\mu$Bq/kg), comparable to the best reported results of low-temperature detectors based on Li$_2$MoO$_4$ with natural or $^{100}$Mo-enriched molybdenum content. Prospects of Li$_2$$^{100\textnormal{depl}}$MoO$_4$ bolometers for use in rare-event search experiments are briefly discussed.}


\begin{document}




\section{Introduction}
\label{sec:intro}

Crystal scintillators are widely used in searches for rare-event processes (like two-$\nu$ and $\nu$-less double-$\beta$ decays, rare $\alpha$ and $\beta$ decays, dark matter particles), particularly using technologies of low-temperature detectors \cite{Pirro:2017,Bellini:2018,Biassoni:2020,Poda:2021,Zolotarova:2021a}. Among them, molybdenum containing crystals represent a long-standing interest for double-$\beta$ decay searches \cite{Pirro:2017,Bellini:2018,Biassoni:2020,Poda:2021,Zolotarova:2021a}, mainly in isotope $^{100}$Mo (possible also for $^{92}$Mo and $^{98}$Mo \cite{Tretyak:2002}); this isotope is also promising for solar and supernova neutrino detection \cite{Ejiri:2000,Ejiri:2002,Ejiri:2014,Ejiri:2017}. 
Different compounds with natural and a few ones with $^{100}$Mo-enriched molybdenum content have been developed and investigated as low-temperature detectors with simultaneous heat and scintillation detection, i.e. scintillating bolometers \cite{Pirro:2017,Bellini:2018,Biassoni:2020,Poda:2021,Zolotarova:2021a}. Lithium molybdate (Li$_{2}$MoO$_4$) is found to be one of the most promising Mo-containing scintillators for such applications \cite{Cardani:2013,Bekker:2016,Armengaud:2017}. Natural and $^{100}$Mo-enriched Li$_{2}$MoO$_4$ (Li$_2$$^{100}$MoO$_4$) have been developed within the LUMINEU project and used in searches for $^{100}$Mo double-$\beta$ decay \cite{Armengaud:2017,Grigorieva:2017,Poda:2017a,Armengaud:2020b}. The LUMINEU technology of scintillating bolometers has been adopted for CUPID-Mo double-$\beta$ experiment \cite{Armengaud:2020a,Armengaud:2021,Augier:2022,Augier:2023a}. Such detector material is also selected for CUPID \cite{CUPIDInterestGroup:2019inu,Armatol:2021a,Armatol:2021b,Alfonso:2022,CrossCupidTower:2023a} and CROSS \cite{Bandac:2020,Bandac:2021,Armatol:2021b,CrossCupidTower:2023a} projects, as well as it represents a great interest for the AMoRE experiment \cite{Alenkov:2015,Kim:2020a,Kim:2020b,Kim:2022a,Kim:2022b}. 

Lithium is also an element with a long-standing interest for rare-event search experiments. The dominant isotope, i.e. $^{7}$Li (92\% in natural lithium), is a light nucleus with non-zero spin, thus a viable candidate to probe spin-dependent dark matter interaction \cite{Bednyakov:2005,Abdelhameed:2019a}.  Moreover, $^{7}$Li is a good target to search for solar axions, via a resonant absorption of an axion by $^{7}$Li and its subsequent $\gamma$ deexcitation \cite{Krcmar:2001,Belli:2008a,Barinova:2010,Belli:2012a,Cardani:2013}. 
Last but not least, presence of $^{6}$Li (8\% in natural lithium) allows neutron detection via a $^{6}$Li(n, t)$\alpha$ reaction, characterized by a high cross-section to thermal neutrons; also, enrichment in $^{6}$Li (up to 95\%) is feasible and it can enhance the detection efficiency. Thus, Li-containing detectors are of special interest for neutron flux monitoring in rare-event search experiments \cite{Martinez:2012,Coron:2016}.

In this work we present development and investigation of scintillating bolometers based on a new type of the Li$_{2}$MoO$_4$ compound, produced from molybdenum depleted in $^{100}$Mo (Li$_2$$^{100\textnormal{depl}}$MoO$_4$). The low concentration of the double-$\beta$ active isotope $^{100}$Mo significantly reduces the related background counting rate, which is on the level of 10 mBq/kg in a Li$_2$$^{100}$MoO$_4$ crystal (an order of magnitude lower for the natural one), and it can be a dominant internal background of this material. Therefore, Li$_2$$^{100\textnormal{depl}}$MoO$_4$ looks more suitable than natural or $^{100}$Mo-enriched Li$_{2}$MoO$_4$ crystals for dark matter and axion search experiments with $^{7}$Li, for double-$\beta$ decay searches in $^{92}$Mo and $^{98}$Mo, as well as for $^{6}$Li-based neutron detection. Moreover, Li$_2$$^{100\textnormal{depl}}$MoO$_4$ can be used in bolometric experiments to search for double-$\beta$ decay in $^{100}$Mo,  as a complementary detector for better understanding a background model. Therefore, a goal of this work is to study prospects of Li$_2$$^{100\textnormal{depl}}$MoO$_4$ low-temperature detectors for rare-event search applications.

\section{Development and test of Li$_2$$^{100\textnormal{depl}}$MoO$_4$ scintillating bolometers}


\subsection{Crystals production and construction of detectors }

In this study we used two Li$_2$$^{100\textnormal{depl}}$MoO$_4$ scintillation crystals with a size of 45$\times$45$\times$45 mm and a mass of around 0.28 kg each, identical to Li$_2$$^{100}$MoO$_4$ crystal produced for the CROSS experiment \cite{Bandac:2020}. The samples were cut from the same crystal boule grown by the low-thermal-gradient Czochralski technique, as detailed in \cite{Grigorieva:2020}. A 4N purity lithium carbonate, selected for the LUMINEU and CUPID-Mo crystals production \cite{Grigorieva:2017,Armengaud:2017,Armengaud:2020a}, and molybdenum oxide depleted in $^{100}$Mo ($\sim$0.01\% of $^{100}$Mo; i.e. 1000 times lower than in natural Mo) have been used as starting materials. The $^{100\textnormal{depl}}$MoO$_3$ material purification, the solid-phase synthesis of the Li$_2$$^{100\textnormal{depl}}$MoO$_4$ compound, and the crystal growth in a Pt crucible using the low-thermal-gradient Czochralski method (double crystalization approach) were realized following protocols of Mo-containing crystals production developed by LUMINEU \cite{Berge:2014,Grigorieva:2017,Armengaud:2017} and adopted by CUPID-Mo \cite{Armengaud:2020a}. A large crystal boule (the cylindrical part is around $\oslash$60$\times$100~mm) has been grown with a crystal yield of about 80\% from the initial charge \cite{Grigorieva:2020} and two twin cubic samples (with a few mm chamfers) were produced.

\begin{figure}
\centering
\includegraphics[width=0.95\textwidth]{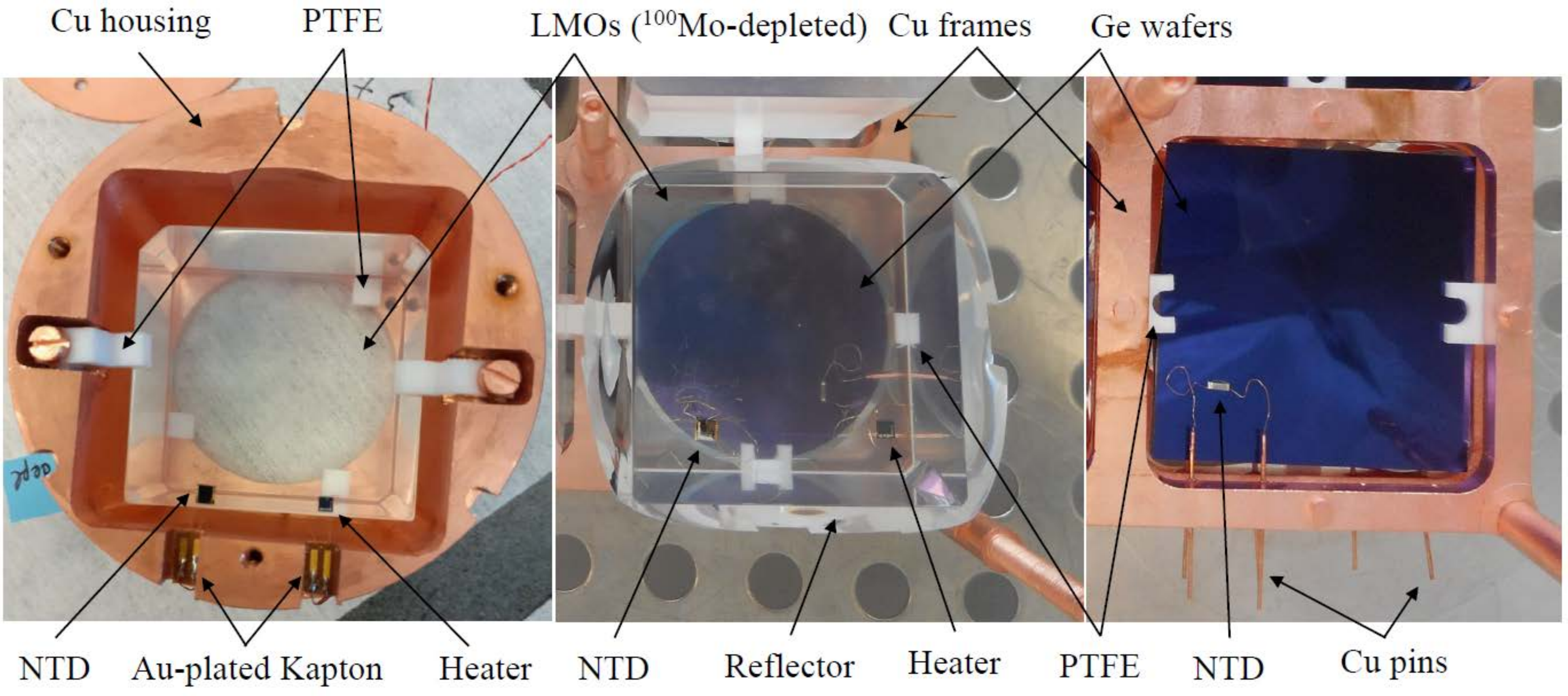}
\caption{Photographs of Li$_2$$^{100\textnormal{depl}}$MoO$_4$ low-temperature detectors LMO-depl-1 (Left) and  LMO-depl-2 (Middle). Both crystals have two epoxy-glued sensors;   the left one is an NTD Ge thermistor, while the right one is a P-doped Si heater. Each scintillator was accompanied by a circular bolometric Ge light detector, as one seen in transparent of the crystal on the middle panel. An additional square-shaped Ge light detector (Right) was used for the LMO-depl-2 sample. All light detectors were instrumented with an NTD Ge sensor.}
\label{fig:Detector_Photo}
\end{figure}

The assembly of detectors have been carried out in an ISO class 4 clean room of the IJClab (Orsay, France). The first Li$_2$$^{100\textnormal{depl}}$MoO$_4$ sample (LMO-depl-1), corresponding to the upper part of the boule, was mounted inside a Cu housing using polytetrafluoroethylene (PTFE) supporting elements and Cu screws, as shown in Figure \ref{fig:Detector_Photo} (Left). The holder design is rather similar to one used in bolometric measurements with the same-size Li$_2$$^{100}$MoO$_4$ \cite{Armatol:2021b} and slightly larger TeO$_2$ \cite{Berge:2018} crystals. 
The second Li$_2$$^{100\textnormal{depl}}$MoO$_4$ sample (LMO-depl-2), corresponding to the bottom part of the boule, was assembled using Cu frames and columns, and PTFE pieces, as seen in Figure \ref{fig:Detector_Photo} (Middle). This sample was a part of a twelve-crystal array, in which all the other crystals are Li$_2$$^{100}$MoO$_4$ \cite{CrossCupidTower:2023a}. 
The lateral side of the LMO-depl-2 crystal was surrounded by a reflective film (Vikuiti{\texttrademark}) to improve the light collection, while the Cu housing of the LMO-depl-1 detector serves as a reflective cavity, however the aperture reduces a direct sight of the light detector. 
Thus, the light collection is non-optimal both due to a poor reflectivity of copper and low scintillation photons collection efficiency on the light detector(s).

To detect particle interactions, each Li$_2$$^{100\textnormal{depl}}$MoO$_4$ crystal was instrumented with a Neutron Transmutation Doped Ge \cite{Haller:1994} thermistor (NTD). A sensor with a size of 3$\times$3$\times$1 mm was epoxy glued onto the crystal's surface using 6 spots of a bi-component glue (Araldite\textsuperscript{\textregistered} Rapid). The temperature-dependent resistance of NTDs can be approximated as $R(T) = R_0 \cdot e^{(T_0/T )^{0.5}}$ with the parameters $R_0$ $\sim$ 1~$\Omega$ and $T_0$ $\sim$ 3.7~K. The NTD sensors with similar irradiation parameters have been used in the CUPID-Mo experiment \cite{Armengaud:2020a}, and CUPID related R\&D tests \cite{Armatol:2021a,Armatol:2021b,Alfonso:2022,CrossCupidTower:2023a}. Also, a  P-doped Si heater \cite{Andreotti:2012} was glued on each crystal with a veil of the epoxy glue. This heating element is exploited to inject by Jules effect power, which can be used e.g. for the stabilization of the thermal gain \cite{Alessandrello:1998}, optimisation of the detector working point, heating of a device if necessary (as in CUPID-0 \cite{Azzolini:2018tum}), pile-up simulations \cite{Armatol:2021}.
To provide electrical contacts, the NTDs and heaters were wire-bonded using $\oslash$25-$\mu$m Au wires.

To allow for the detection of Li$_2$$^{100\textnormal{depl}}$MoO$_4$ scintillation, we accompanied crystals with bolometric detectors based on electronic-grade purity Ge wafers, supplied by Umicore (Belgium) \cite{Umicore}. Two of them have a circular shape (with a size of $\oslash$45$\times$0.18 mm each), while the third device is square-shaped  (45$\times$45$\times$0.30 mm). All the Ge disks are coated on both sides with a 70 nm SiO layer aiming at reducing the light reflection \cite{Mancuso:2014,Azzolini:2018tum,Armengaud:2020a}. Smaller NTD sensors (3$\times$1$\times$1~mm or 3$\times$0.7$\times$1~mm) are attached to the Ge wafers with a veil of the epoxy glue. The Ge disks are  PTFE-clamped in the Cu structure. A single circular light detector (LD-1-c) has been coupled to the LMO-depl-1 crystal, while both circular (LD-2-c) and square-shaped (LD-2-s) bolometric photodetectors viewed the LMO-depl-2 crystal. The mounted Ge light detectors are shown in Figure \ref{fig:Detector_Photo} (Middle and Right).


\subsection{Operation at Canfranc underground laboratory}

\begin{figure}[hbt]
\centering
\includegraphics[width=0.95\textwidth]{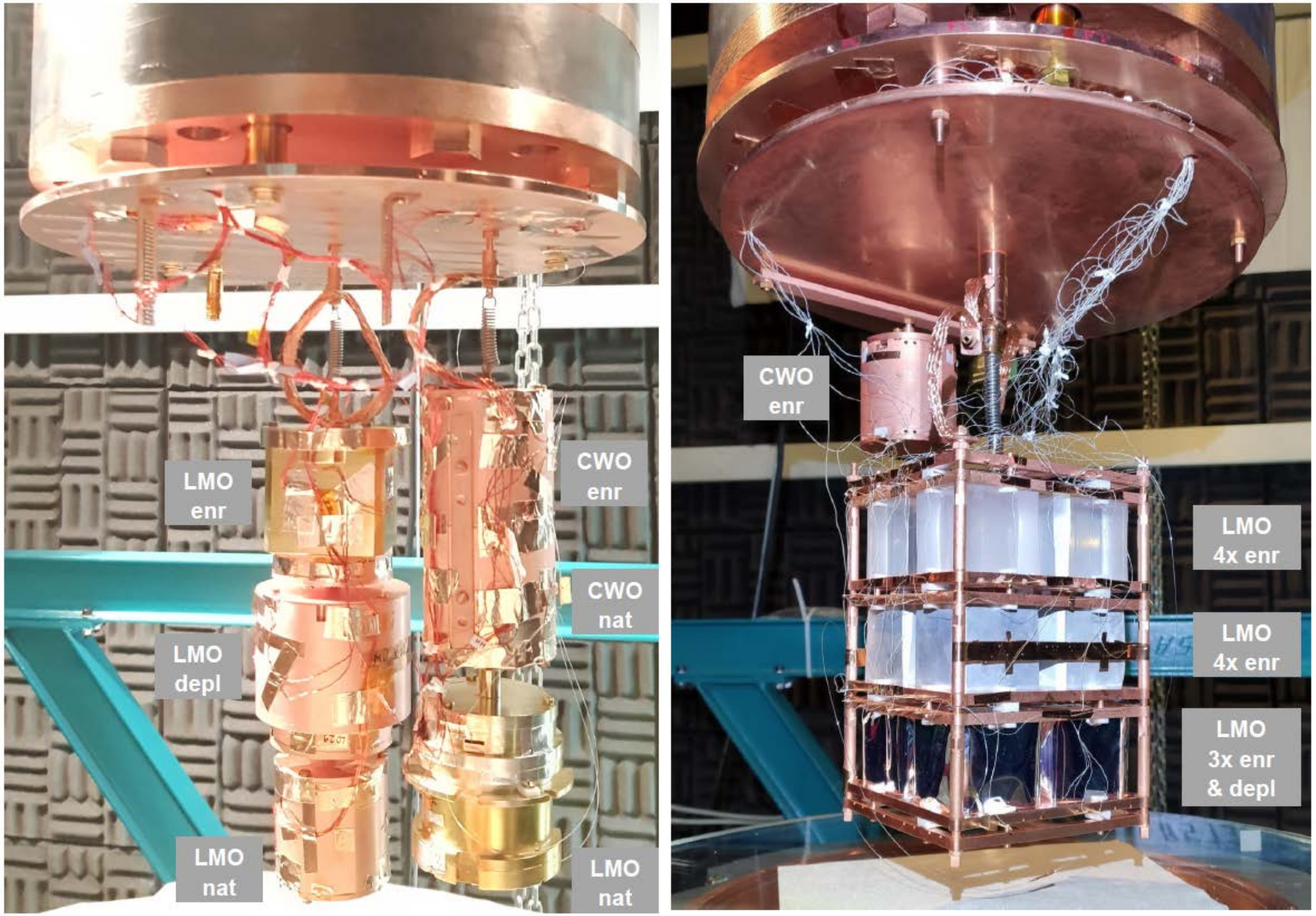}
\caption{Detector configurations in the C2U cryogenic runs at the LSC, in which Li$_2$$^{100\textnormal{depl}}$MoO$_4$ scintillating bolometers LMO-depl-1 (Left) and LMO-depl-2 (Right) were operated. Other scintillating bolometers are based on Li$_2$MoO$_4$ crystals with natural (CROSS \cite{Khalife:2021} and CLYMENE \cite{Velazquez:2017,Stelian:2020} R\&D) and $^{100}$Mo-enriched (joint CROSS and CUPID R\&D \cite{Armatol:2021b,CrossCupidTower:2023a}) molybdenum content, as well as on natural and $^{116}$Cd-enriched CdWO$_4$ crystals \cite{Olivieri:2020,Zolotarova:2020,Helis:2020,Helis:2021}.}
\label{fig:Detector_lsc}
\end{figure}

The Li$_2$$^{100\textnormal{depl}}$MoO$_4$ scintillating bolometers were operated in the CROSS cryogenic set-up (C2U) \cite{Olivieri:2020,Armatol:2021b} at the Canfranc underground laboratory (LSC, Spain), providing a substantial reduction of the cosmic muon flux thanks to the rock overburden \cite{Trzaska:2019}. The detectors were assembled as parts of scintillating bolometer arrays and installed inside the cryostat, as illustrated in Figure \ref{fig:Detector_lsc}. The facility exploits the HEXA-DRY dilution fridge by CryoConcept (France), which is equipped by the Ultra-Quiet Technology{\texttrademark} \cite{UQT} to decouple a pulse tube (Cryomech PT415) from the dilution unit, thus reducing vibrations \cite{Olivieri:2017}. To further improve noise conditions, the detector arrays were spring-suspended from the cold plate of the cryostat. The detector volume inside the cryostat is shielded on the top by a 13~cm thick disk made of interleaved lead and copper (partially seen in Figure \ref{fig:Detector_lsc}), while the outer vacuum chamber is surrounded by a 25~cm thick layer of low-radioactivity lead. In addition, a deradonized air flow around the cryostat had been supplied during the whole time of the experiment with the LMO-depl-2 detector.

After reaching the base temperature of the cryostat ($\sim$10~mK), we regulated the detector plate temperature at 18 and then 12~mK for measurements with the LMO-depl-1 bolometer, and at 14~mK for the LMO-depl-2 operation. The control and readout of bolometers was done with the help of a low-noise room-temperature DC front-end electronics, restyled from the Cuoricino experiment \cite{Arnaboldi:2002}. The data acquisition (DAQ) system was composed of two 12-channel boards with integrated 24-bit ADC and a programmable 6-pole Bessel-Thomson anti-aliasing filter (the cut-off frequency of the low-pass filter was set at 300~Hz) \cite{Carniti:2020,Carniti:2023}. 

In order to find an optimal working point of the detectors, representing the best signal-to-noise ratio, we spanned the bolometric response with respect to the current across an NTD \cite{Novati:2019}. We used heaters to inject thermal pulses to both Li$_2$$^{100\textnormal{depl}}$MoO$_4$ bolometers, while LED generated photons, transmitted from a room-temperature LED (the emission maximum is at $\sim$880~nm) through an optic fibre, were exploited for the light detectors. The heater/LED signal injection was performed with the help of a wave-function generator (Keysight 33500B). As a result of the optimization, we polarized NTDs of the detectors at a few nA current, reducing NTD resistances from hundreds M$\Omega$ (at low power) to a few M$\Omega$ (at the working point).

For each operational temperature we performed measurements with a removable $^{232}$Th $\gamma$ source, made of a thoriated tungsten wire, and without the source (data are referred to calibration and background respectively). Even if the $\gamma$ source is primarily conceived for calibration of the Li$_2$$^{100\textnormal{depl}}$MoO$_4$ bolometers, we also used it to evaluate the energy scale of the light detectors similar to \cite{Berge:2018,Armengaud:2020a}. 

The continuous data of each channel were acquired with a sampling rate of 2~kS/s and stored on a disk for the offline analysis. We processed data with the help of a MATLAB-based analysis tool \cite{Mancuso:2016}, which implements the signal processing using the Gatti-Manfredi optimum filter \cite{Gatti:1986} to maximize the signal-to-noise ratio. In order to apply the filter, we used data-based information about the signal shape (represented by an average signal of tens of high-energy events) and the measured noise (represented by 10000 waveforms with no signal). The data were triggered with a threshold corresponding to 5$\sigma$ of the filtered noise. For each triggered signal, we collected an information about its amplitudes (i.e. energy) and several pulse-shape parameters. The results of detectors characterization are presented in the next section.

\section{Characterization of Li$_2$$^{100\textnormal{depl}}$MoO$_4$ scintillating bolometers}

\subsection{Performance of detectors}

At first, we investigated the recorded bolometric signals in terms of time constants of the signal shape. The rising part of a signal is commonly characterized by the rise time parameter, which is computed as the time required by the signal to increase from 10\% to 90\% of its amplitude. The descending part is described by the decay time, defined here as the time required to drop from 90\% to 30\% of signal amplitude. The rise and decay time parameters of the operated low-temperature detectors are summarized in Table \ref{tab:LMO_performance}.
We find that the Li$_2$$^{100\textnormal{depl}}$MoO$_4$ bolometers have signals with the rise time of $\sim$20 ms and the decay time $\sim$100 ms. These time constants are similar to the values reported for NTD-instrumented low-temperature detectors based on similar-size 
Li$_2$MoO$_4$ crystals produced from molybdenum with the natural isotopic abundance and from enriched in $^{100}$Mo \cite{Bekker:2016,Armengaud:2017,Armengaud:2020a,Armatol:2021a,Armatol:2021b,CrossCupidTower:2023a}. 
The Ge light detectors, being gram-scale bolometric devices equipped with smaller NTDs (i.e. reduced heat capacity compared to the Li$_2$$^{100\textnormal{depl}}$MoO$_4$ bolometers), have an order of magnitude faster response, typical for such devices \cite{Beeman:2013b,Armengaud:2017,Azzolini:2018tum,Armengaud:2020a,Armatol:2021a,Armatol:2021b,CrossCupidTower:2023a}.


\begin{table}
 \caption{Performance of Li$_2$$^{100\textnormal{depl}}$MoO$_4$ scintillating bolometers and Ge light detectors. We report the detector plate temperature, the NTD resistance, the signal rise and decay time parameters, the detector sensitivity, the baseline noise resolution, and the energy resolution at a given energy. }
\footnotesize
\begin{center}
\begin{tabular}{l|c|c|c|c|c|c|l}
 \hline
Bolometer & Tempe- & Resistance & Rise  & Decay & Sensitivity & FWHM$_{Noise}$ & FWHM (keV) \\
~ & rature & of NTD  & time & time & (nV/keV) & (keV) & at energy (keV)  \\
~ & (mK) & (M$\Omega$) & (ms) & (ms) & ~ & ~ & ~  \\
 \hline
LMO-depl-1   & 18 & 2.4  & 16    & 112    & 17    & 3.66(3)   & 5.9(10) at 1765 \\
~            & 12 & 6.5  & 20    & 115    & 37    & 2.18(3)   & 5.8(3) at 2615 \\
LMO-depl-2   & 14 & 3.0  & 16    & 97     & 29    & 3.80(3)   & 6.8(3) at 2615 \\
\hline
LD-1-c       & 18 & 1.6  & 1.5   & 9.0   & 1200   & 0.097(1)  & 0.174(4) at 5.9 \\
~            & 12 & 2.6  & 1.6   & 10.5  & 1380   & 0.100(1)  & 0.146(2) at 5.9 \\
LD-2-s       & 14 & 0.47 & 1.6   & 5.2   & 380    & 0.343(5)  & 0.94(6) at 17.5 \\
LD-2-c       & 14 & 4.4  & 2.0   & 7.8   & 2200   & 0.059(1)  & 0.90(6) at 17.5 \\
 \hline
 \end{tabular}
  \label{tab:LMO_performance}
 \end{center}
 \end{table}

\normalsize

\begin{figure}[hbt]
\centering
\includegraphics[width=0.49\textwidth]{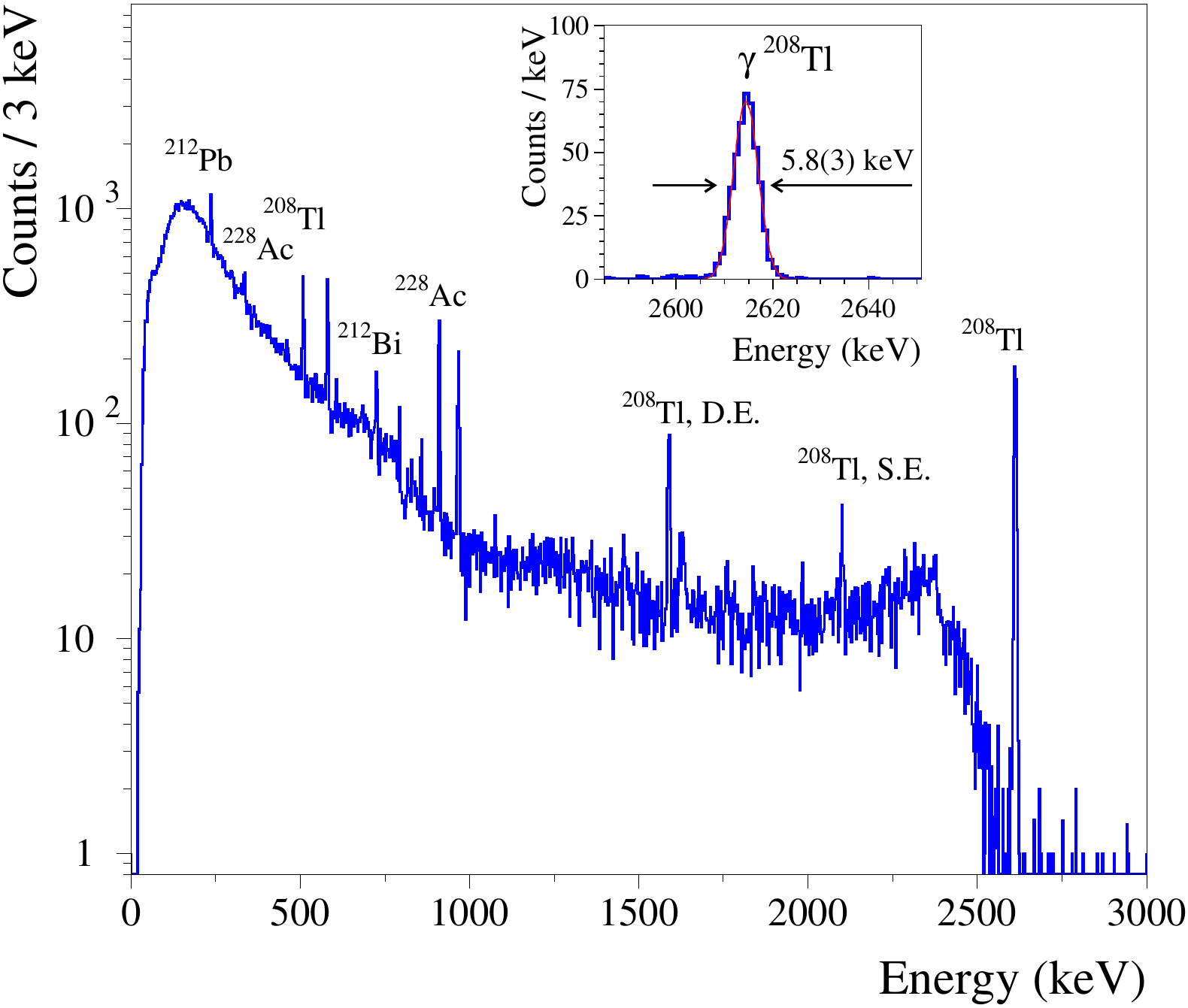}
\includegraphics[width=0.48\textwidth]{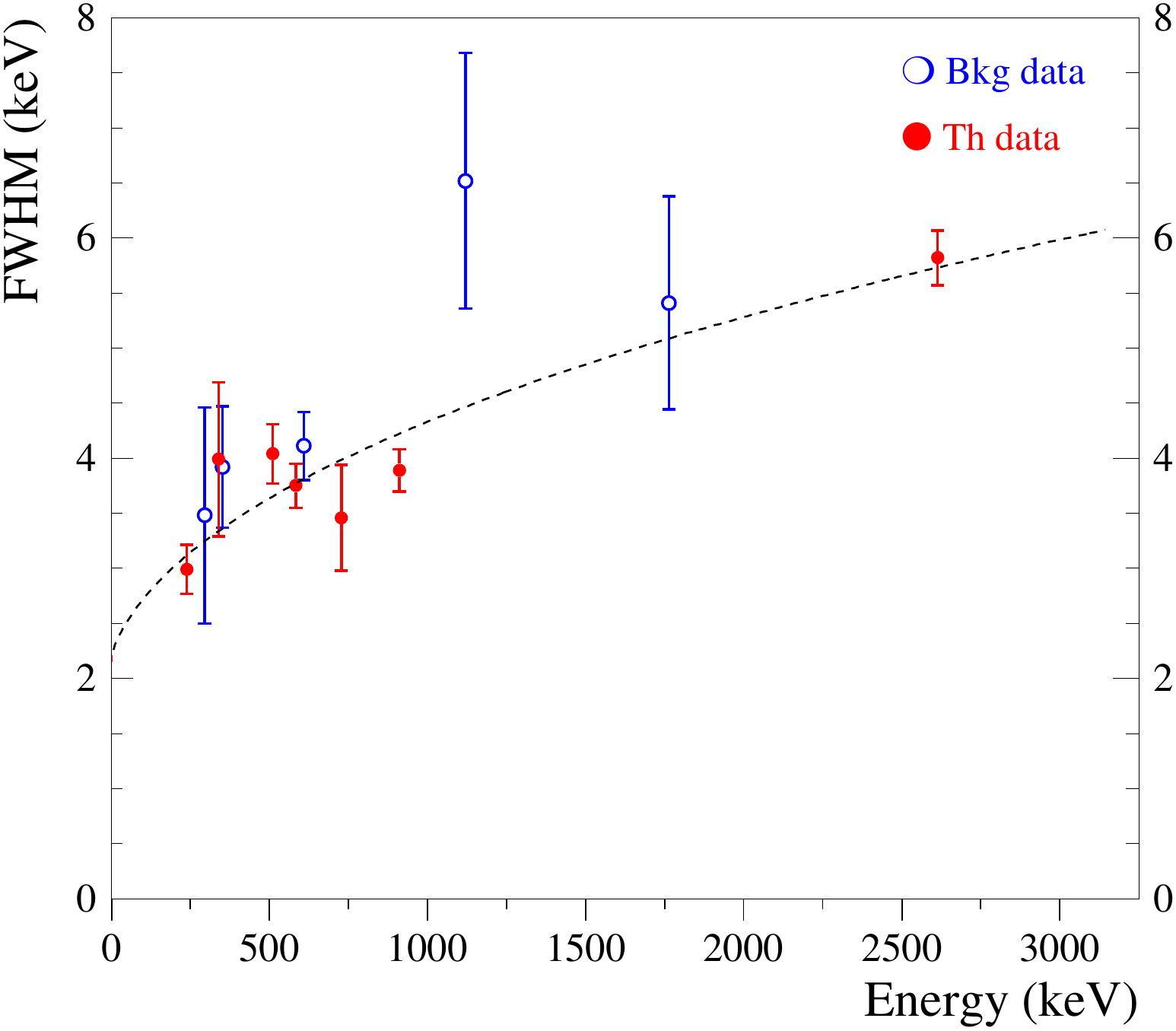}
\caption{(Left) Energy spectrum of a $^{232}$Th source measured with the Li$_2$$^{100\textnormal{depl}}$MoO$_4$ (LMO-depl-1) bolometer, operated at 12 mK over 125 h. The most intense $\gamma$-ray peaks observed in the spectrum are labeled by their origin; D.E. and S.E. mean double and single escape peaks, respectively. A fit to the 2615 keV peak of $^{208}$Tl is shown in the inset; the energy resolution is 5.8(3) keV FWHM. (Right) The energy dependence of the Li$_2$$^{100\textnormal{depl}}$MoO$_4$ (LMO-depl-1) bolometer energy resolution in calibration (red, 125 h) and background (blue, 1109 h) data acquired at 12 mK. The fit is shown by a dotted line.}
  \label{fig:LMO_Th_spectrum}
\end{figure}

Then, using an amplitude distribution of events recorded in the calibration runs, we calibrated the energy scale of bolometers and evaluated their sensitivity, expressed as a voltage amplitude per unit of a deposited energy (e.g. nV/keV), as well as their energy resolution in the limit of zero amplitude (baseline noise) and at a mono-energetic radiation. In order to calibrate the Li$_2$$^{100\textnormal{depl}}$MoO$_4$ bolometers, we used the most intense $\gamma$ quanta emitted by the $^{232}$Th source in the energy interval of 0.2--2.6 MeV, as illustrated in Figure \ref{fig:LMO_Th_spectrum} (Left). In the background data we also relied on the presence of $\gamma$ peaks from environmental radioactivity (mainly $\gamma$-active daughters of radon, $^{214}$Pb and $^{214}$Bi, examples are given below). The sensitivities of the Li$_2$$^{100\textnormal{depl}}$MoO$_4$ bolometers were measured as $\sim$20--40~nV/keV (see in Table \ref{tab:LMO_performance}); the LMO-depl-1 signal increases by a factor 2 at a colder temperature of the heat sink. Taking into account that the chosen working points are characterised by relatively high NTD currents, the achieved sensitivities are not exceptional among Li$_2$MoO$_4$-based bolometers (the highest reported values are $\sim$100--150 nV/keV) \cite{Bekker:2016,Armengaud:2017,Armengaud:2020a,Armatol:2021a,Armatol:2021b,CrossCupidTower:2023a}. Also, the baseline noise is found to be rather low, $\sim$2--4~keV FWHM (Table \ref{tab:LMO_performance}), and similar to early reported performance of NTD-instrumented Li$_2$MoO$_4$ bolometers (the best performing detectors have $\sim$1 keV FWHM noise) \cite{Bekker:2016,Armengaud:2017,Armengaud:2020a,Armatol:2021a,Armatol:2021b,CrossCupidTower:2023a}. To further improve the baseline noise (e.g. for dark matter search applications), one can reduce the absorber's volume (i.e. heat capacity) and/or use an advanced performance phonon sensor technology \cite{Abdelhameed:2019a}.
Despite of not extraordinary noise resolution, both Li$_2$$^{100\textnormal{depl}}$MoO$_4$ bolometers show a comparatively high energy resolution as exampled in Figure \ref{fig:LMO_Th_spectrum} (Right). As it is also seen in Table \ref{tab:LMO_performance}, the energy resolution for high-energy $\gamma$ quanta (1.8 and 2.6 MeV) is only a factor 2--3 worse than the resolution at zero energy, which is a good feature of Li$_2$MoO$_4$ bolometers \cite{Armengaud:2017}. Consequently, the Li$_2$$^{100\textnormal{depl}}$MoO$_4$ energy resolution at the 2615 keV MeV $\gamma$'s, listed in Table \ref{tab:LMO_performance} and illustrated in Figure \ref{fig:LMO_Th_spectrum} (Left, Inset), is among the best reported for Li$_2$MoO$_4$ low-temperature detectors \cite{Armengaud:2017,Armengaud:2020a,Armatol:2021a,Armatol:2021b,Alfonso:2022,CrossCupidTower:2023a}.

\begin{figure}[hbt]
\centering
\includegraphics[width=0.49\textwidth]{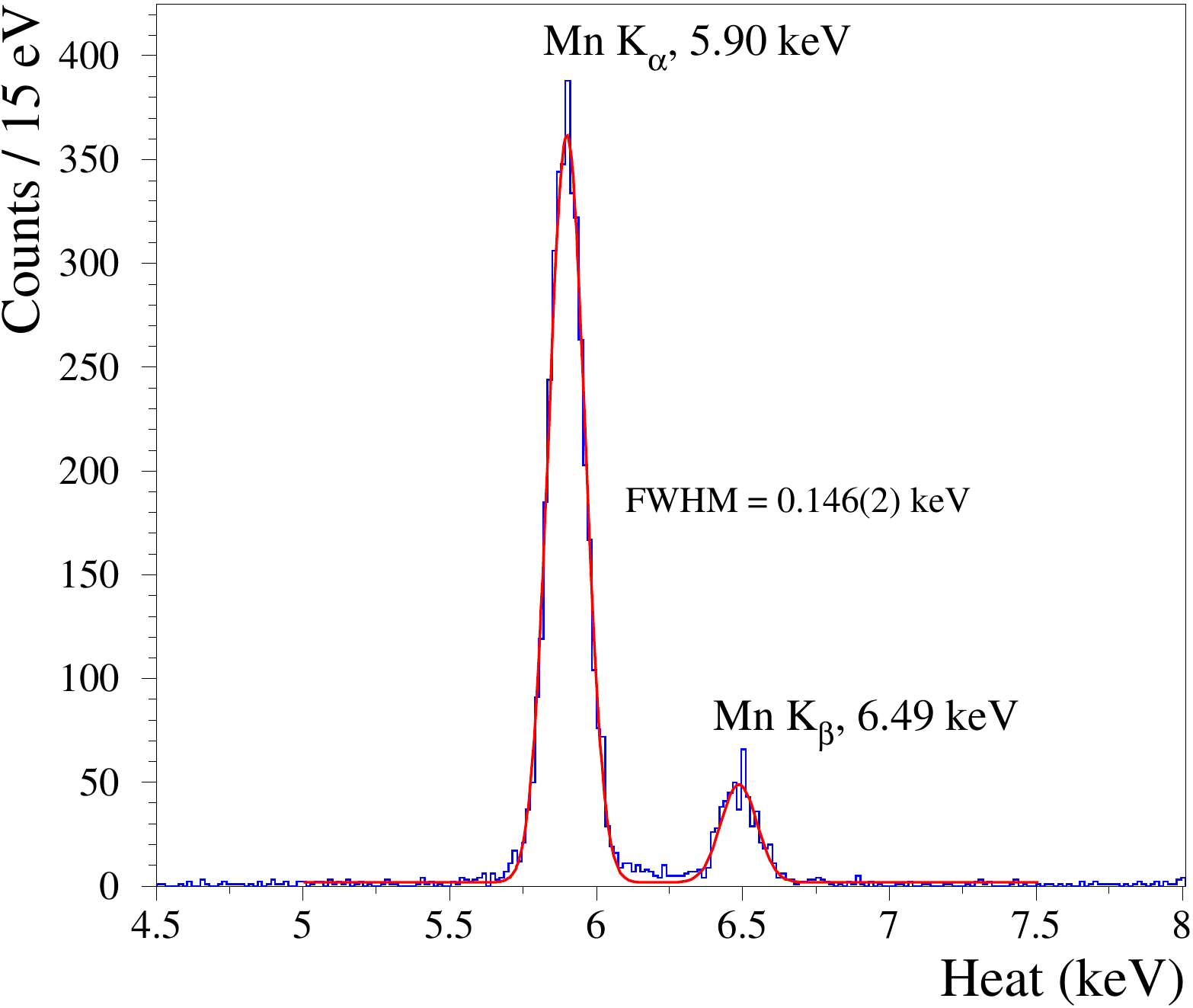}
\includegraphics[width=0.49\textwidth]{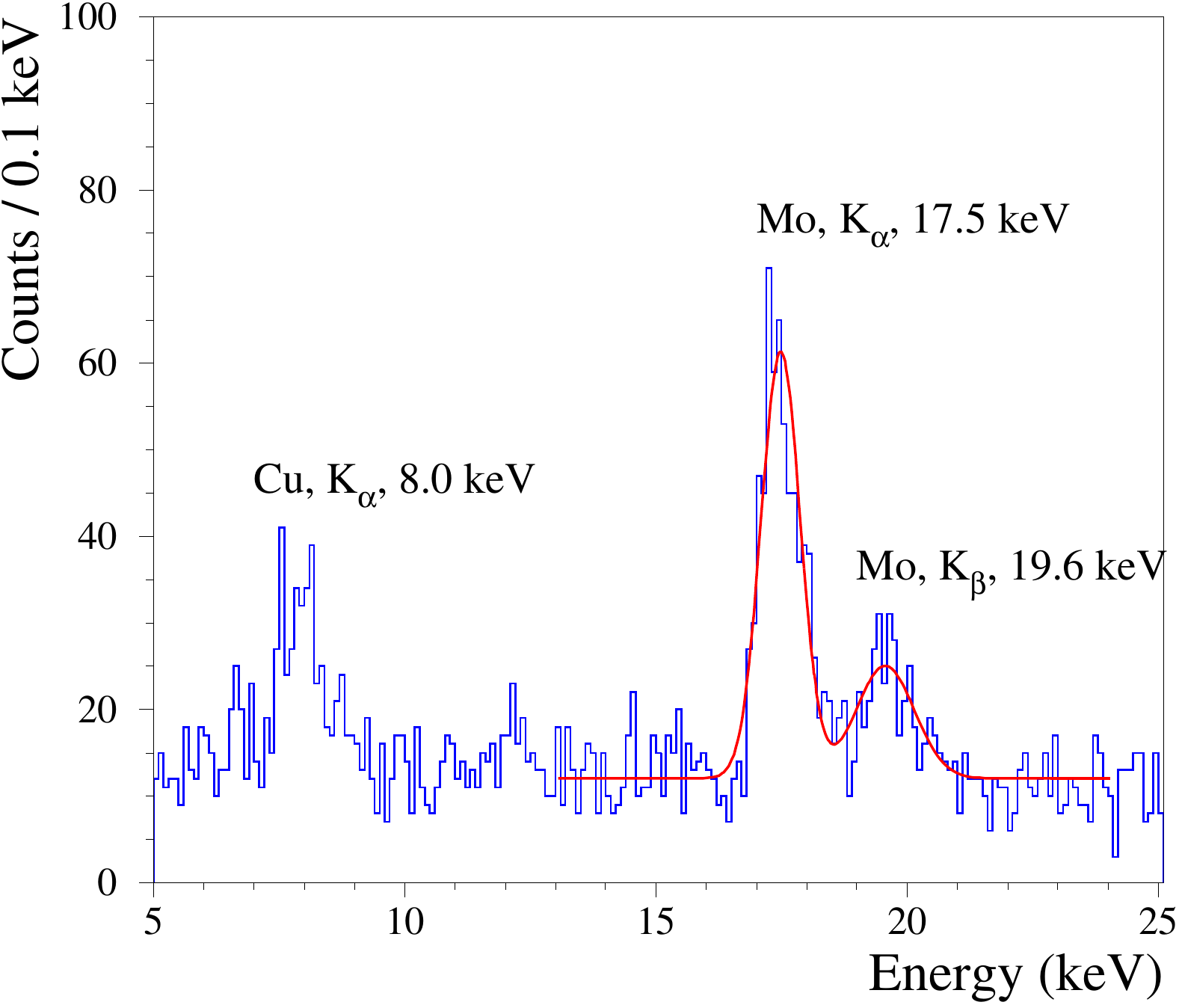}
\caption{Energy spectra of X-rays from a close $^{55}$Fe X-ray source (Left) and Cu / Mo X-rays induced by the $^{232}$Th $\gamma$-ray source (Right), measured by the LD-1-c (1109 h of background data) and LD-2-c (266 h, calibration) bolometers respectively. Fits to the spectra with two Gaussians and a linear background component are shown by solid red lines.}
  \label{fig:LD_spectra}
\end{figure}

Aiming a permanent calibration during measurements, the LD-1-c was supplied by an $^{55}$Fe X-ray source, which irradiated a Ge side opposite to the LMO-depl-1 crystal. This source emits a doublet of Mn K$_{\alpha}$ and K$_{\beta}$ X-rays with energies of 5.9 and 6.5 keV, and intensities 25\% and 3\%, respectively; an example of the energy spectrum is presented in  Figure~\ref{fig:LD_spectra} (Left). To overcome the absence of permanent X-ray sources in the assembly of the LMO-depl-2 scintillating bolometer, we irradiated this detector with the $^{232}$Th $\gamma$ source to induce X-ray fluorescence of the materials close to the light detectors, i.e. in the Cu structure and in the crystal. An illustration of the resulting spectrum is shown in  Figure~\ref{fig:LD_spectra} (Right). So, knowing the energy scale, we observed good sensitivity of two light detectors (1.2--2.2 $\mu$V/keV), while the third device had a reduced value (0.4 $\mu$V/keV) due to stronger polarization of NTD, as exhibited by lower NTD resistance (see Table \ref{tab:LMO_performance}). This performance is typical for such type of bolometric detectors with NTD thermistors; a further gain is also feasible by reducing heat capacity of the sensor / absorber (e.g. see \cite{Armengaud:2017} and references therein) and/or upgrading with Neganov-Trofimov-Luke-effect-based signal amplification \cite{Novati:2019}. The less sensitive light detector had comparatively modest noise resolution (about 300 eV FWHM), while the other two detectors demonstrated a rather low noise, 60--100~eV FWHM (e.g. see in \cite{Poda:2021}). The resolution of the 6 keV X-ray peak is found to be close to the baseline noise value, while a more broader 17 keV Mo X-ray peak has been detected by both light detectors irrespective of 5 times difference in the baseline noise.  This effect can be explained by a position-dependent response of such thin bolometers \cite{Armengaud:2017}.

\subsection{Scintillation detection and particle identification}

\begin{figure}[hbt]
\centering
\includegraphics[width=0.49\textwidth]{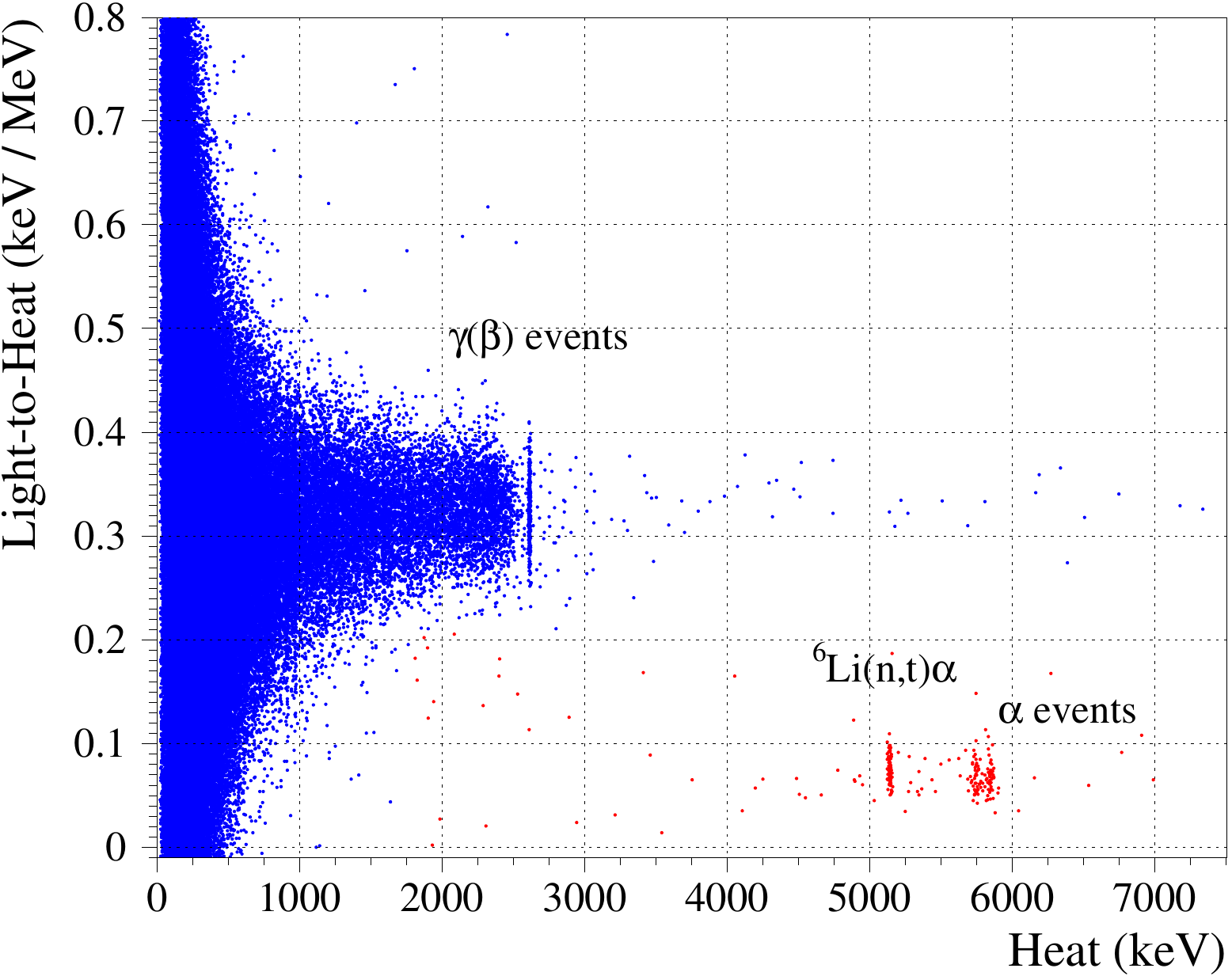}
\includegraphics[width=0.49\textwidth]{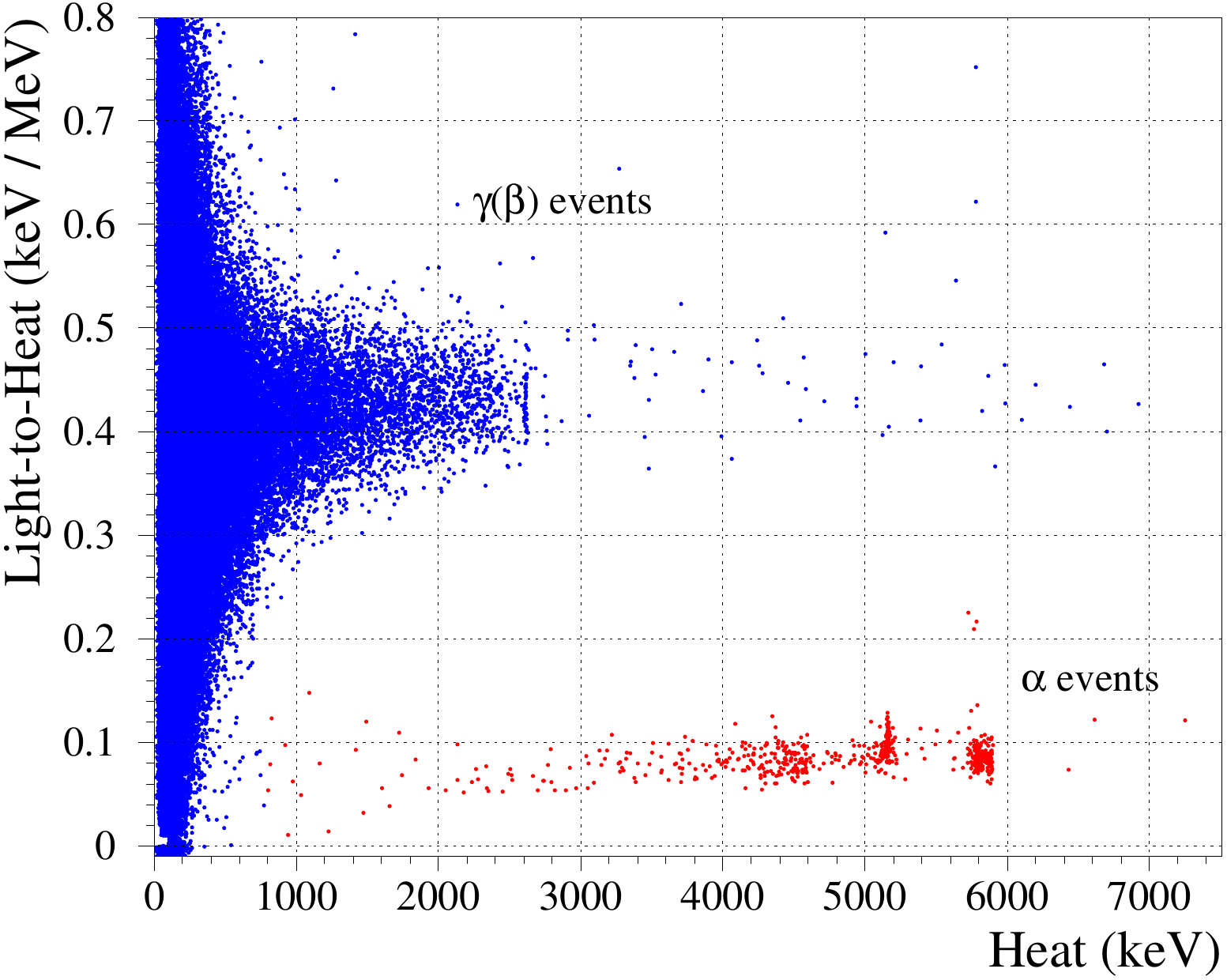}
\includegraphics[width=0.49\textwidth]{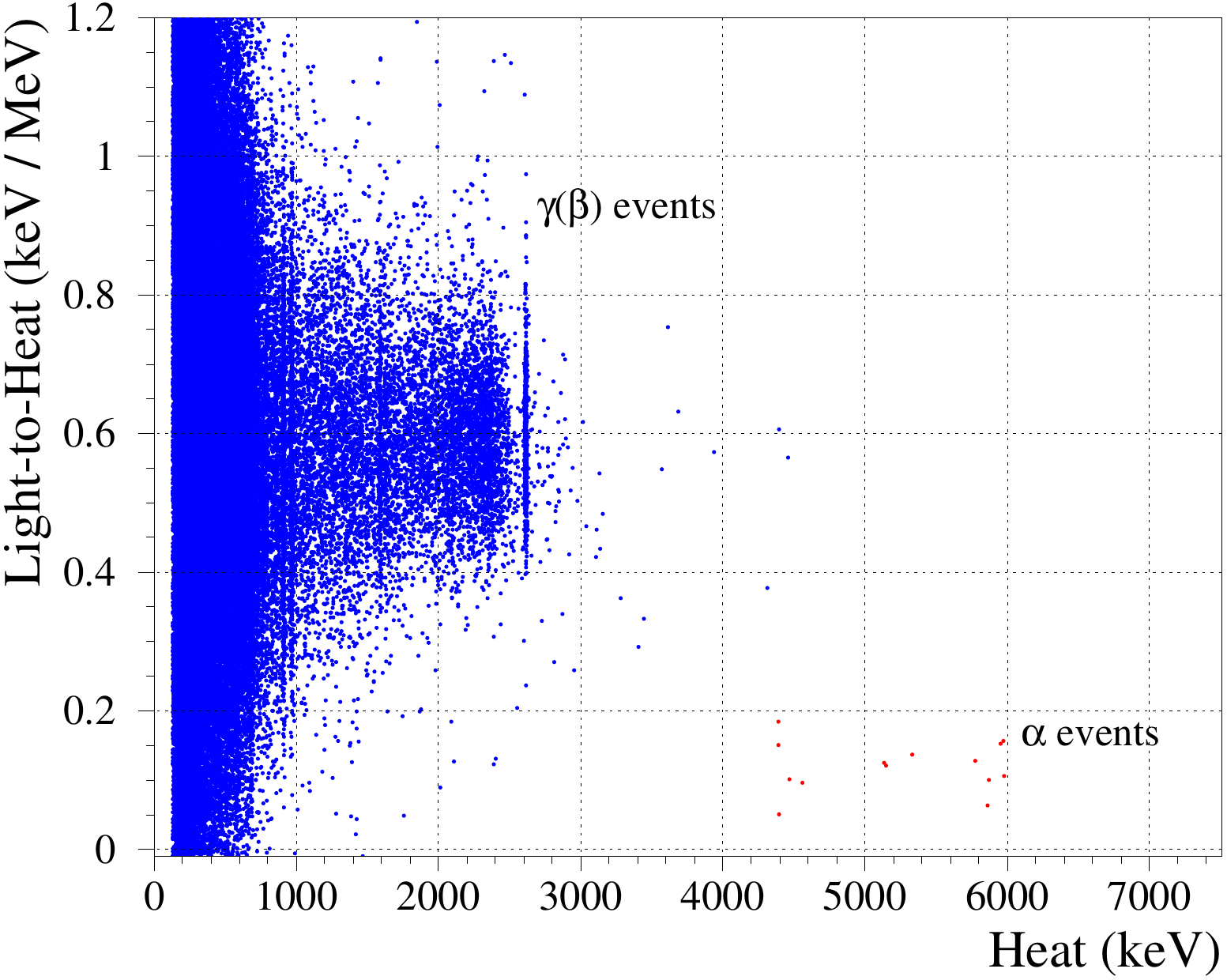}
\caption{Scintillation (light-to-heat parameter) versus heat energy release measured by the Li$_2$$^{100\textnormal{depl}}$MoO$_4$ scintillating bolometers. Top panel, left shows a sum of calibration (125 h) and background (1109 h) data of the LMO-depl-1. The LMO-depl-2 events detected in background (top, right; 1536~h) and calibration (bottom; 111~h) measurements in coincidences with the bolometric photodetectors LD-2-c and LD-2-s, respectively.}
\label{fig:LMO_LY-vs-Heat}
\end{figure}

A combination of both heat and scintillation channels of a scintillating bolometer can provide particle identification, exploiting  the dependence of the light output on the energy loss mechanism (i.e. particle type) \cite{Pirro:2005ar,Poda:2021}. In order to find coincidences between signals in the Li$_2$$^{100\textnormal{depl}}$MoO$_4$ bolometers and in the associated light detectors, the latter channels were processed using trigger positions of Li$_2$$^{100\textnormal{depl}}$MoO$_4$ events and accounting for a difference in time response (see Table \ref{tab:LMO_performance}), similarly to \cite{Piperno:2011}. It is convenient to present such data by normalizing the light detector signal on the corresponding heat energy release, the so-called light-to-heat parameter (L/H) in units of keV/MeV. The dependence of the L/H parameter on energy and type of particles impinged the Li$_2$$^{100\textnormal{depl}}$MoO$_4$ detectors is illustrated in Figure \ref{fig:LMO_LY-vs-Heat}.

Several populations of events are clearly seen in each data presented in Figure \ref{fig:LMO_LY-vs-Heat}. The most dominant one, distributed mainly below 3 MeV, is originated by $\gamma$($\beta$) particles. We selected $\gamma(\beta)$'s with energies above 2~MeV to evaluate the corresponding light-to-heat parameter (L/H$_{\gamma(\beta)}$), reported in Table \ref{tab:LMO_LY}. The lowest L/H$_{\gamma(\beta)}$ value ($\sim$0.3 keV/MeV) is obtained for the LMO-depl-1 and it is expected due to sub-optimal scintillation light collection conditions (e.g. the aperture between the crystal and the photodetector, the Cu surrounding instead of a reflective foil, smaller light-detector area). Indeed, the twin detector (LMO-depl-2), being surrounded by the reflective foil and coupled to the same-size photodetector, detected about 30\% more scintillation energy ($\sim$0.4 keV/MeV), while the square-shaped light detector allowed almost to double the scintillation signal ($\sim$0.6 keV/MeV). 
At higher energies, above $\sim$3 MeV, we see populations of events, which are characterised by scintillation reduced to $\sim$20\% compared to $\gamma$($\beta$)'s, as seen in Table \ref{tab:LMO_LY}. These events are originated by $\alpha$'s from  either U/Th traces of detector bulk / surface contamination or a U source and to $\alpha$+t particles, products of neutron capture on $^6$Li. 
Despite the different light collection conditions the characteristics of the Li$_2$$^{100depl}$MoO$_4$ scintillating bolometers, $L/H_{\gamma(\beta)}$ and $QF_{\alpha}$ listed in Table \ref{tab:LMO_LY}, are similar to the ones reported for detectors based on Li$_2$MoO$_4$ crystals produced from molybdenum with the natural isotopic composition and  enriched in $^{100}$Mo \cite{Armengaud:2017,Armengaud:2020a,Poda:2020,Armatol:2021a,Armatol:2021b,Alfonso:2022}.

\begin{table}
 \caption{Results on scintillation detection with Li$_2$$^{100\textnormal{depl}}$MoO$_4$ scintillating bolometers. We report the light-to-heat ratios for $\gamma$($\beta$) events (L/H$_{\gamma(\beta)}$) and the quenching factors for scintillation induced by $\alpha$'s of $^{210}$Po (QF$_{\alpha}$). }
\footnotesize
\begin{center}
\begin{tabular}{c|c|c|c}
 \hline
Crystal & Photodetector & L/H$_{\gamma(\beta)}$ & QF$_{\alpha}$  \\
~ & ~ & (keV/MeV) & ~    \\
 \hline
LMO-depl-1   & LD-1-c  & 0.33(3)    & 0.21(4)      \\

\hline
LMO-depl-2   & LD-2-c  & 0.44(3)    & 0.19(4)        \\
~            & LD-2-s  & 0.59(9)    & ~      \\
 \hline
 \end{tabular}
  \label{tab:LMO_LY}
 \end{center}
 \end{table}

\normalsize

\subsection{Radiopurity of Li$_2$$^{100\textnormal{depl}}$MoO$_4$ crystals}

\begin{figure}[hbt]
\centering
\includegraphics[width=0.49\textwidth]{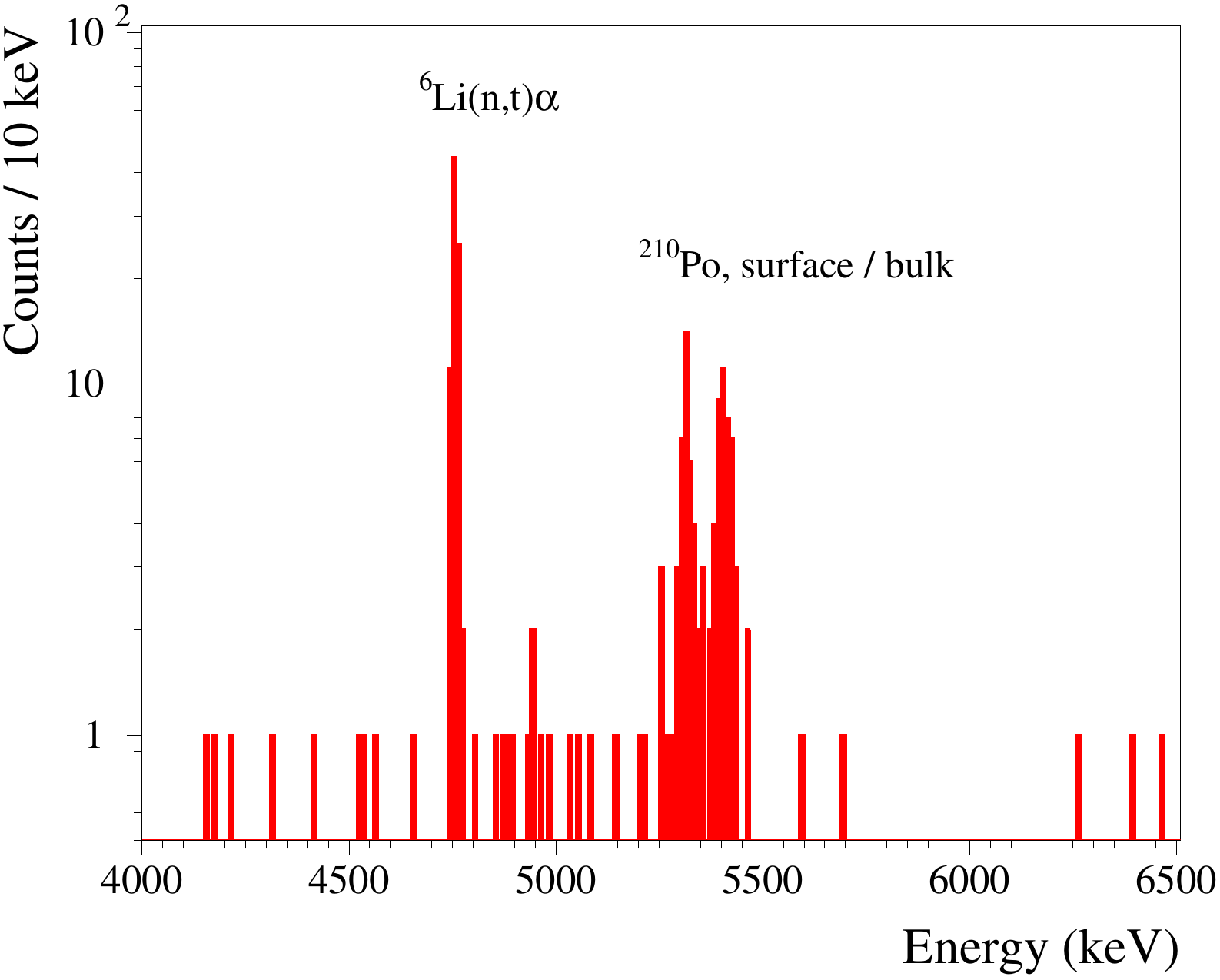}
\includegraphics[width=0.49\textwidth]{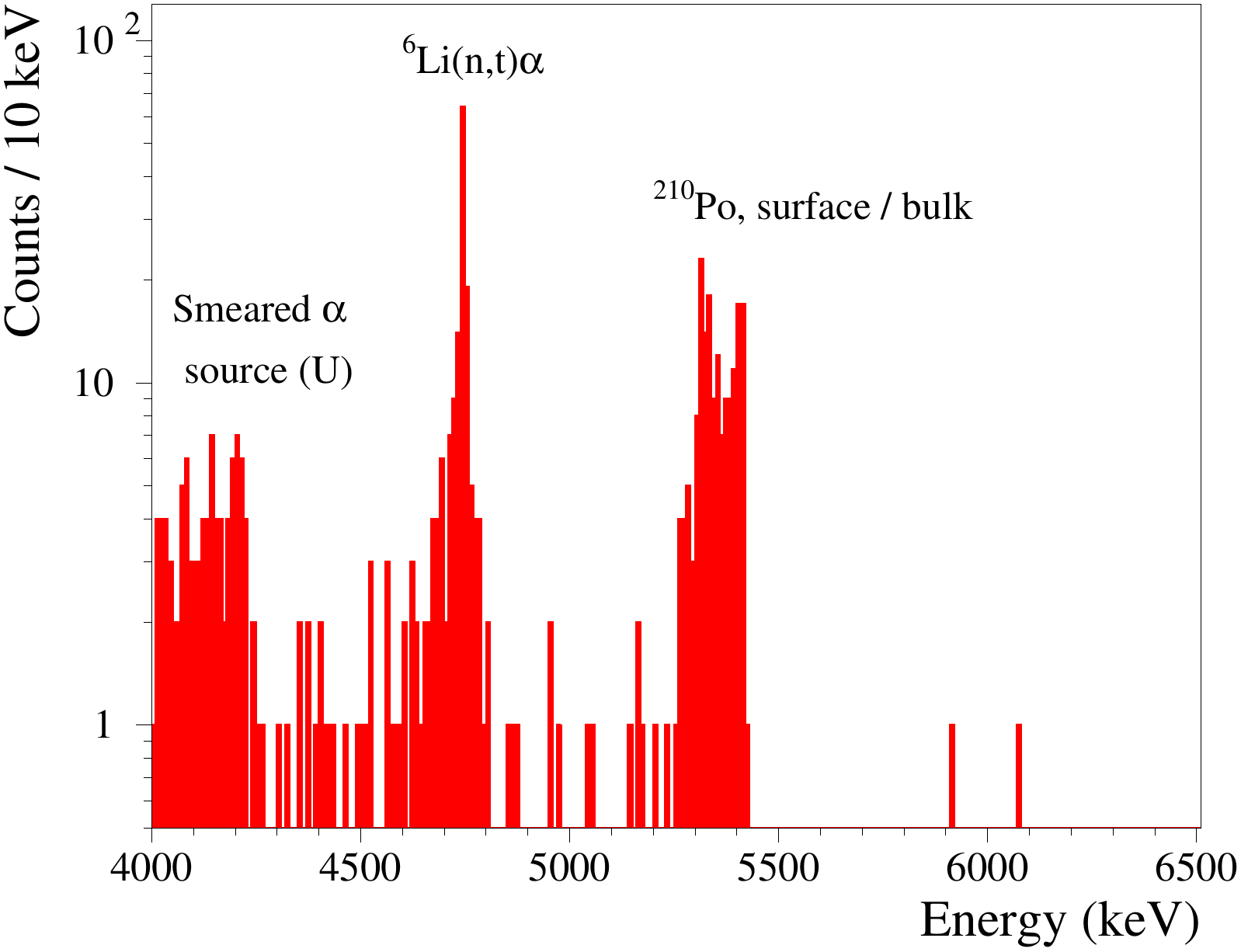}
\caption{Energy spectra of $\alpha$ events detected by Li$_2$$^{100\textnormal{depl}}$MoO$_4$ scintillating bolometers made of crystals LMO-depl-1 (Left; 1109~h of measurements) and LMO-depl-2 (Right; 1528~h), operated underground in the C2U facility.}
\label{fig:LMO_Bkg_alpha}
\end{figure}

Thanks to efficient particle identification (Figure \ref{fig:LMO_LY-vs-Heat}) and comparatively long background measurements, we can investigate radiopurity of the Li$_2$$^{100\textnormal{depl}}$MoO$_4$ crystals with a high sensitivity to $\alpha$-active radionuclides of the U/Th decay chains. With this aim, we selected $\alpha$ particles from the data of both detectors and re-calibrated spectra to alpha-energy scale for analysis of $\alpha$ contaminants, the resulting data are shown in Figure \ref{fig:LMO_Bkg_alpha}.

It is seen in Figure \ref{fig:LMO_Bkg_alpha} that the $\alpha$ spectra of the Li$_2$$^{100\textnormal{depl}}$MoO$_4$ crystals are rather similar with the exception of the energy region below $\sim$4.7~MeV, which is populated by $\alpha$ particles from a $^{238}$U/$^{234}$U source used in the set-up nearby the LMO-depl-2 detector. The spectra contain only two peak-like structures, which is a clear indication of a high internal radiopurity. Moreover, the first peak at $\sim$4.8~MeV is originated by the detection of products ($\alpha$ plus triton) of thermal neutron captures by $^6$Li; this peak is detected by both bolometers with a similar rate of 1.8(2) counts/day. Furthermore, a doublet of 5.3 and 5.4 MeV peaks is a summed contribution of detector surface and crystal bulk contaminations by $^{210}$Po; in the later case $\alpha$ particle and $^{206}$Pb nuclear recoil (taking away 0.1 MeV energy)  were detected. The activity of $^{210}$Po in both crystals is the same $\sim$35~$\mu$Bq/kg; it is originated by a residual $^{210}$Pb contamination, typical for scintillators, particularly for Mo-containing compounds  \cite{Armengaud:2015,Armengaud:2017,Armengaud:2020a,Danevich:2017,Danevich:2018}.

We found no clear evidence of other $\alpha$-active radionuclides from U/Th chains in the $\alpha$ spectra of the Li$_2$$^{100\textnormal{depl}}$MoO$_4$ detectors and thus we set upper limits on their activities in the crystal bulk, using the Feldman-Cousins approach \cite{Feldman:1998}. As a signal, we took all events in the region $\pm$25~keV around a $Q$-value of a radionuclide searched for, while background estimate is done in the neighbour energy interval out of the $Q$-values of U/Th $\alpha$'s, as e.g. detailed in \cite{Armengaud:2015}. The results of the study of Li$_2$$^{100\textnormal{depl}}$MoO$_4$ crystals radiopurity are summarized in Table \ref{tab:LMO_radiopurity}. The limits on U/Th activity are obtained on the level of a few $\mu$Bq/kg; therefore radiopurity of the two Li$_2$$^{100\textnormal{depl}}$MoO$_4$ samples is similar to the purity level of Li$_2$$^{100}$MoO$_4$ crystals produced for LUMINEU \cite{Armengaud:2017,Poda:2017a}, CUPID-Mo \cite{Armengaud:2020a,Poda:2020} and CROSS \cite{Armatol:2021b,CrossCupidTower:2023a} experiments, thanks to the same purification and crystallization protocols applied.

\begin{table}
 \caption{Radioactive contamination of the Li$_2$$^{100\textnormal{depl}}$MoO$_4$ crystals by $\alpha$-active radionuclides from $^{238}$U/$^{232}$Th families (their $Q$-values are listed in keV). The uncertainties are given at 68\% C.L., while the limits are set at 90\% C.L.}
\footnotesize
\begin{center}
\begin{tabular}{c|c|c|c|c|c|c}
 \hline
Crystal & \multicolumn{6}{c}{Activity ($\mu$Bq/kg)}  \\
\cline{2-7}
~ & $^{232}$Th & $^{228}$Th & $^{238}$U & $^{234}$U & $^{226}$Ra & $^{210}$Po    \\
~ & [4082] & [5520] & [4270] & [4858] & [4871] & [5407] \\
 \hline
LMO-depl-1   & < 2 & < 2  & < 2 & < 5 & < 7    & 35(6)      \\

\hline
LMO-depl-2   & ~ & < 2  & ~ & ~ & < 4    & 36(5)        \\

 \hline
 \end{tabular}
  \label{tab:LMO_radiopurity}
 \end{center}
 \end{table}

\normalsize


\subsection{Background reconstruction capability of Li$_2$$^{100\textnormal{depl}}$MoO$_4$ bolometers}

\begin{figure}[hbt]
\centering
\includegraphics[width=0.7\textwidth]{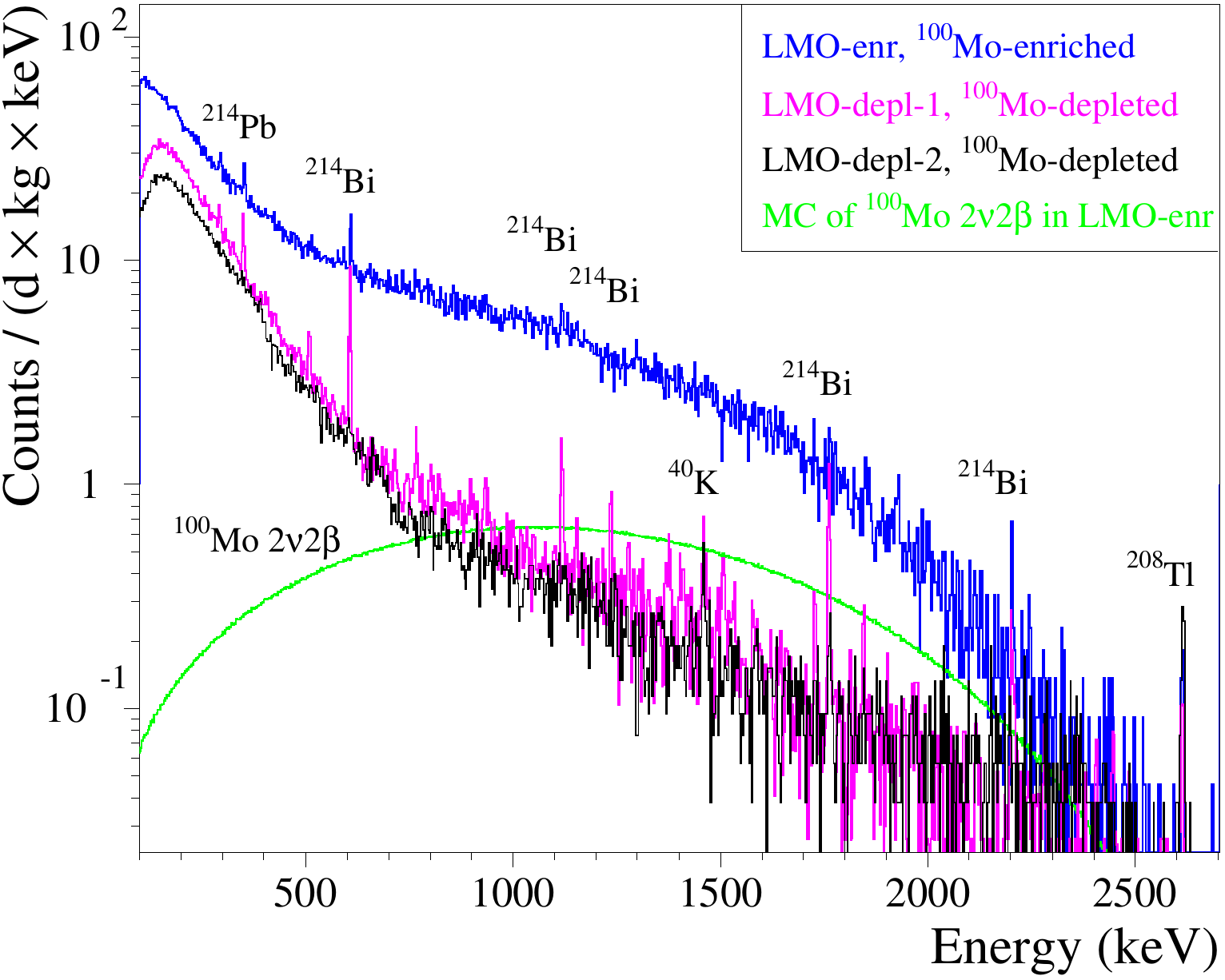}
\caption{Energy spectra of $\gamma$($\beta$) events detected by bolometers based on 0.28 kg lithium molybdate crystals produced from molybdenum either depleted in $^{100}$Mo (pink, LMO-depl-1, and black, LMO-depl-2) or enriched in $^{100}$Mo (blue, LMO-enr data from \cite{Armatol:2021b}), and operated in the C2U set-up at the Canfranc underground laboratory. The detectors LMO-depl-1 and LMO-enr were running together (see Figure \ref{fig:Detector_lsc}), while the LMO-depl-2 detector was measured in the next cryogenic run, in which the set-up was flushed with deradonized air. It is worth noting the LMO-enr bolometer was irradiated by a close $^{238}$U/$^{234}$U $\alpha$ source, which emits $\beta$ particles too, $^{234}$Th ($Q_{\beta}$ = 0.27~MeV) and $^{234m}$Pa ($Q_{\beta}$ = 2.2~MeV). Thus, the difference between the acquired spectra is mainly explained by both the $^{100}$Mo two-$\nu$ double-$\beta$ activity ($\sim$3~mHz, $Q_{\beta\beta}$ = 3.0~MeV) and the $\alpha$-source-induced $\beta$ background (about 20~mHz of $^{234}$Th and $^{234m}$Pa) of the Li$_2$$^{100}$MoO$_4$ detector. A Monte Carlo distribution of the $^{100}$Mo two-$\nu$ double-$\beta$ decay events (green) \cite{Armengaud:2020b} is shown for comparison.}
\label{fig:LMO_Bkg_gamma}
\end{figure}

As mentioned in Section \ref{sec:intro}, $^{100}$Mo represents a great interest for double-$\beta$ decay studies. However, the two-$\nu$ double-$\beta$ decay of $^{100}$Mo ---characterized by a relatively ``fast'' half-life ($\sim$7$\times$10$^{18}$~yr \cite{Barabash:2020,Armengaud:2020b})--- is an important source of background in $\nu$-less double-$\beta$ decay search experiments. Indeed, it generates a 10~mHz rate in a 1 kg $^{100}$Mo-enriched lithium molybdate crystal being a dominant background component in a wide energy interval \cite{Armengaud:2017,Augier:2022,CUPIDMoBkgModel:2023}. But even crystals with natural Mo content have not negligible internal activity of $^{100}$Mo ($\sim$1~mBq/kg). The impact of $^{100}$Mo radioactivity can be seen in Figure \ref{fig:LMO_Bkg_gamma}, where the energy spectra accumulated with $^{100}$Mo-enriched/depleted Li$_2$MoO$_4$ bolometers in a common measurement at the C2U facility are shown. It is worth noting that the background of the Li$_2$$^{100}$MoO$_4$ bolometer was spoiled by a $\beta$-component of the used external $\alpha$ source with a notable activity, around an order of magnitude higher than that of the $^{100}$Mo two-$\nu$ double-$\beta$ decay, as evident in Figure \ref{fig:LMO_Bkg_gamma}. A clear $\gamma$ background reduction exhibited by the LMO-depl-2 bolometer, in comparison to the LMO-depl-1 data, was achieved thanks to using a deradonized air flow around the cryostat shielding. Both Li$_2$$^{100\textnormal{depl}}$MoO$_4$ detectors have similar high background below 0.7 MeV, which is explained by a $^{210}$Bi activity induced by the $^{210}$Pb contamination of the lead shield. Also, the residual $\gamma$($\beta$) activity inside the experimental volume of the set-up, detected by both Li$_2$$^{100\textnormal{depl}}$MoO$_4$ bolometers above $\sim$1~MeV is higher than e.g. in CUPID-Mo \cite{Augier:2022,CUPIDMoBkgModel:2023}, CUPID-0 \cite{Azzolini:2019nmi,Azzolini:2022}, and CUORE \cite{Adams:2021,Adams:2022}. Thus, the difference between the $^{100}$Mo two-$\nu$ double-$\beta$ decay distribution and the background data of the Li$_2$$^{100\textnormal{depl}}$MoO$_4$ detectors, acquired in not fully optimized background conditions, seen in Figure \ref{fig:LMO_Bkg_gamma} is not remarkable as it might be in a better shielded set-up.

It is evident on Figure \ref{fig:LMO_Bkg_gamma} that the Li$_2$$^{100\textnormal{depl}}$MoO$_4$ bolometer allows significantly improved reconstruction of $\gamma$ background, including low-intensity contributions, compared to the Li$_2$$^{100}$MoO$_4$ detector having a dominant double-$\beta$ (and $\beta$) decay events continuum. Therefore, Li$_2$$^{100\textnormal{depl}}$MoO$_4$ low-temperature detectors with high spectrometric performance and high radiopurity can provide a complementary information about a background model in double-$\beta$ decay searches with Li$_2$$^{100}$MoO$_4$ bolometers. A similar example is the CUPID-0 experiment with 2 natural and 24 $^{82}$Se-enriched zinc selenide bolometers \cite{Azzolini:2018tum}), however the detectors from selenium with the natural isotopic concentration were not included to the background model analysis \cite{Azzolini:2019nmi}. Moreover, a combination of Li$_2$$^{100\textnormal{depl}}$MoO$_4$ and Li$_2$$^{100}$MoO$_4$ bolometers of similar performance and purity would allow extracting a half-life and a spectral shape of the $^{100}$Mo two-$\nu$ double-$\beta$ decay. 
A similar approach has been used in several double-$\beta$ searches, e.g. with $^{40}$Ca-/$^{48}$Ca-enriched calcium fluoride scintillation detectors \cite{Mateosian:1966}, a $^{78}$Kr-enriched/depleted gas filled proportional counter \cite{Gavrilyuk:2013}, and a $^{136}$Xe-enriched/depleted xenon gas based time-projection chamber \cite{Novella:2022}. However, the enrichment in $^{48}$Ca, present in natural calcium at a $\sim$0.2\% level, is not yet available in large quantity as for $^{100}$Mo, $^{78}$Kr and $^{136}$Xe \cite{Giuliani:2012a}. Moreover, the very long half-lives of $^{78}$Kr (10$^{22}$~yr) and $^{136}$Xe (10$^{21}$~yr) require a huge exposure to collect a reasonably high statistics of double-$\beta$ decay events, while more than two orders of magnitude faster rate of $^{100}$Mo is an advantage for such study. 

An order of magnitude lower background of the Li$_2$$^{100\textnormal{depl}}$MoO$_4$ bolometer at $\sim$0.5 MeV, compared to the Li$_2$$^{100}$MoO$_4$ one, would provide a higher sensitivity to a peak expected at 478 keV in case of a resonant absorption of solar axions to the first excited state of $^7$Li and its subsequent $\gamma$ deexcitation \cite{Belli:2012a,Cardani:2013}. At lower energies, the contribution of $^{100}$Mo two-$\nu$ double-$\beta$ decay becomes negligible \cite{Armatol:2021b}, making no preference on $^{100}$Mo content in Li$_2$MoO$_4$ crystals for a search of spin-dependent dark matter interaction on $^7$Li. A the same time the presence of $^6$Li allows neutron detection (illustrated above in Figures \ref{fig:LMO_LY-vs-Heat} and \ref{fig:LMO_Bkg_alpha}), which can be exploited for neutron flux monitoring, particularly relevant for dark matter search applications.

\section{Conclusions}

In the present study we demonstrate that scintillating bolometers made of lithium molybdate crystals produced from molybdenum depleted in $^{100}$Mo (Li$_2$$^{100\textnormal{depl}}$MoO$_4$) show high performance, comparable to the devices based on crystals from molybdenum of the natural isotopic composition or enriched in $^{100}$Mo. Thanks to the strict purification and crystallization protocols, developed and already applied to high-sensitivity searches for $^{100}$Mo $\nu$-less double-$\beta$ decay, radiopurity of Li$_2$$^{100\textnormal{depl}}$MoO$_4$ crystals is rather high and compatible to $^{100}$Mo-enriched crystals of the same production line. Thus, given an availability of molybdenum depleted in $^{100}$Mo (a by product of industrial enrichment in $^{100}$Mo) for Li$_2$$^{100\textnormal{depl}}$MoO$_4$ crystals production, as well as high spectrometric performance, efficient scintillation-assisted particle identification capability and high material radiopurity, Li$_2$$^{100\textnormal{depl}}$MoO$_4$ scintillating bolometers show a high potential for applications in rare-event search experiments. In particular such detectors represent a great interest for studies of $^{100}$Mo two-neutrino double-$\beta$ decay, to searches for $^7$Li axions and spin-depend interactions of dark matter particles on $^7$Li, as well as for $^6$Li-based neutron spectroscopy and $\gamma$($\beta$) background control measurements in low-temperature low-background experiments.

\vspace{6pt} 




\funding{``This research received no external funding''}

\acknowledgments{
This work is supported by the European Commission (Project CROSS, Grant No. ERC-2016-ADG, ID 742345). 
This work was also supported by the National Research Foundation of Ukraine under Grant No. 2020.02/0011 and by the National Academy of Sciences of Ukraine in the framework of the project ``Development of bolometric experiments for the search for double beta decay'', the grant number 0121U111684. 
Y.A.B., V.D.G., E.P.M., and V.N.S. were supported by the Ministry of Science and Higher Education of the Russian Federation N121031700314-5. 
Russian and Ukrainian scientists have given and give crucial contributions to CROSS. For this reason, the CROSS collaboration is particularly sensitive to the current situation in Ukraine. The position of the collaboration leadership on this matter, approved by majority, is expressed at \href{https://a2c.ijclab.in2p3.fr/en/a2c-home-en/assd-home-en/assd-cross/}{https://a2c.ijclab.in2p3.fr/en/a2c-home-en/assd-home-en/assd-cross/}. Majority of the work described here was completed before February 24, 2022.
}

\conflictsofinterest{``The authors declare no conflict of interest.''} 




\appendixtitles{no} 


\reftitle{References}


\begin{thebibliography}{-------}
\providecommand{\natexlab}[1]{#1}

\bibitem[Pirro and Mauskopf(2017)]{Pirro:2017}
Pirro, S.; Mauskopf, P.
\newblock {Advances in Bolometer Technology for Fundamental Physics}.
\newblock {\em Annu. Rev. Nucl. Part. Sci.} {\bf 2017}, {\em 67},~161.

\bibitem[Bellini(2018)]{Bellini:2018}
Bellini, F.
\newblock Potentialities of the future technical improvements in the search of
  rare nuclear decays by bolometers.
\newblock {\em Int. J. Mod. Phys. A} {\bf 2018}, {\em 33},~1843003.

\bibitem[Biassoni and Cremonesi(2020)]{Biassoni:2020}
Biassoni, M.; Cremonesi, O.
\newblock {Search for neutrino-less double beta decay with thermal detectors}.
\newblock {\em Prog. Part. Nucl. Phys.} {\bf 2020}, {\em 114},~103803.

\bibitem[Poda(2021)]{Poda:2021}
Poda, D.
\newblock {Scintillation in Low-Temperature Particle Detectors}.
\newblock {\em Physics} {\bf 2021}, {\em 3},~473--535.

\bibitem[Zolotarova(2021)]{Zolotarova:2021a}
Zolotarova, A.
\newblock Bolometric Double Beta Decay Experiments: Review and Prospects.
\newblock {\em Symmetry} {\bf 2021}, {\em 13},~2255.

\bibitem[Tretyak and Zdesenko(2002)]{Tretyak:2002}
Tretyak, V.I.; Zdesenko, Y.G.
\newblock {Tables of double beta decay data --- an update}.
\newblock {\em At. Data Nucl. Data Tables} {\bf 2002}, {\em 80},~83--116.

\bibitem[Ejiri \em{et~al.}(2000)Ejiri, Engel, Hazama, Krastev, Kudomi, and
  Robertson]{Ejiri:2000}
Ejiri, H.; Engel, J.; Hazama, R.; Krastev, P.; Kudomi, N.; Robertson, R.
\newblock {Spectroscopy of Double-Beta and Inverse-Beta Decays from $^{100}$Mo
  for Neutrinos}.
\newblock {\em Phys. Rev. Lett.} {\bf 2000}, {\em 85},~2917.

\bibitem[Ejiri \em{et~al.}(2002)Ejiri, Engel, and Kudomi]{Ejiri:2002}
Ejiri, H.; Engel, J.; Kudomi, N.
\newblock {Supernova-neutrino studies with $^{100}$Mo}.
\newblock {\em Phys. Lett. B} {\bf 2002}, {\em 530},~27--32.

\bibitem[Ejiri and Elliott(2014)]{Ejiri:2014}
Ejiri, H.; Elliott, S.R.
\newblock {Charged current neutrino cross section for solar neutrinos, and
  background to $\beta\beta$($0\nu$) experiments}.
\newblock {\em Phys. Rev. C} {\bf 2014}, {\em 89},~055501.

\bibitem[Ejiri and Elliott(2017)]{Ejiri:2017}
Ejiri, H.; Elliott, S.R.
\newblock {Solar neutrino interactions with the double-$\beta$ decay nuclei
  $^{82}$Se, $^{100}$Mo, and $^{150}$Nd}.
\newblock {\em Phys. Rev. C} {\bf 2017}, {\em 95},~055501.

\bibitem[Cardani \em{et~al.}(2013)Cardani, Casali, Nagorny, Pattavina, Piperno,
  Barinova, Beeman, Bellini, Danevich, Di~Domizio, et~al.]{Cardani:2013}
Cardani, L.; Casali, N.; Nagorny, S.; Pattavina, L.; Piperno, G.; Barinova, O.;
  Beeman, J.; Bellini, F.; Danevich, F.; Di~Domizio, S.; others.
\newblock {Development of a Li$_2$MoO$_4$ scintillating bolometer for low
  background physics}.
\newblock {\em J. Instrum.} {\bf 2013}, {\em 8},~P10002.

\bibitem[Bekker \em{et~al.}(2016)Bekker, Coron, Danevich, Degoda, Giuliani,
  Grigorieva, Ivannikova, Mancuso, De~Marcillac, Moroz, et~al.]{Bekker:2016}
Bekker, T.; Coron, N.; Danevich, F.; Degoda, V.Y.; Giuliani, A.; Grigorieva,
  V.; Ivannikova, N.; Mancuso, M.; De~Marcillac, P.; Moroz, I.; others.
\newblock {Aboveground test of an advanced Li$_2$MoO$_4$ scintillating
  bolometer to search for neutrinoless double beta decay of $^{100}$Mo}.
\newblock {\em Astropart. Phys.} {\bf 2016}, {\em 72},~38.

\bibitem[Armengaud \em{et~al.}(2017)Armengaud, Augier, Barabash, Beeman,
  Bekker, Bellini, Beno{\^\i}t, Berg{\'e}, Bergmann, Billard,
  et~al.]{Armengaud:2017}
Armengaud, E.; Augier, C.; Barabash, A.; Beeman, J.; Bekker, T.; Bellini, F.;
  Beno{\^\i}t, A.; Berg{\'e}, L.; Bergmann, T.; Billard, J.; others.
\newblock {Development of $^{100}$Mo-containing scintillating bolometers for a
  high-sensitivity neutrinoless double-beta decay search}.
\newblock {\em Eur. Phys. J. C} {\bf 2017}, {\em 77},~785.

\bibitem[Grigorieva \em{et~al.}(2017)Grigorieva, Shlegel, Bekker, Ivannikova,
  Giuliani, de~Marcillac, Marnieros, Novati, Olivieri, Poda, Nones, Zolotarova,
  and Danevich]{Grigorieva:2017}
Grigorieva, V.D.; Shlegel, V.; Bekker, T.; Ivannikova, N.; Giuliani, A.;
  de~Marcillac, P.; Marnieros, S.; Novati, V.; Olivieri, E.; Poda, D.; Nones,
  C.; Zolotarova, A.; Danevich, F.
\newblock {Li$_2$MoO$_4$ crystals grown by low thermal gradient Czochralski
  technique}.
\newblock {\em J. Mat. Sci. Eng. B} {\bf 2017}, {\em 7},~63.

\bibitem[Poda and {LUMINEU, EDELWEISS, CUPID-0/Mo
  collaborations}(2017)]{Poda:2017a}
Poda, D.V.; {LUMINEU, EDELWEISS, CUPID-0/Mo collaborations}.
\newblock {$^{100}$Mo-enriched Li$_2$MoO$_4$ scintillating bolometers for $0\nu
  2\beta$ decay search: from LUMINEU to CUPID-0/Mo projects}.
\newblock {\em AIP Conf. Proc.} {\bf 2017}, {\em 1894},~020017.

\bibitem[Armengaud \em{et~al.}(2020{\natexlab{a}})Armengaud, Augier, Barabash,
  Bellini, Benato, Beno{\^\i}t, Beretta, Berg{\'e}, Billard, Borovlev,
  et~al.]{Armengaud:2020b}
Armengaud, E.; Augier, C.; Barabash, A.; Bellini, F.; Benato, G.; Beno{\^\i}t,
  A.; Beretta, M.; Berg{\'e}, L.; Billard, J.; Borovlev, Y.A.; others.
\newblock {Precise measurement of $2\nu\beta\beta$ decay of $^{100}$Mo with the
  CUPID-Mo detection technology}.
\newblock {\em Eur. Phys. J. C} {\bf 2020}, {\em 80},~674.

\bibitem[Armengaud \em{et~al.}(2020{\natexlab{b}})Armengaud, Augier, Barabash,
  Bellini, Benato, Beno{\^\i}t, Beretta, Berg{\'e}, Billard, Borovlev,
  et~al.]{Armengaud:2020a}
Armengaud, E.; Augier, C.; Barabash, A.; Bellini, F.; Benato, G.; Beno{\^\i}t,
  A.; Beretta, M.; Berg{\'e}, L.; Billard, J.; Borovlev, Y.A.; others.
\newblock {The CUPID-Mo experiment for neutrinoless double-beta decay:
  performance and prospects}.
\newblock {\em Eur. Phys. J. C} {\bf 2020}, {\em 80},~44.

\bibitem[Armengaud \em{et~al.}(2021)Armengaud, Augier, Barabash, Bellini,
  Benato, Benoit, Beretta, Berg{\'e}, Billard, Borovlev,
  et~al.]{Armengaud:2021}
Armengaud, E.; Augier, C.; Barabash, A.; Bellini, F.; Benato, G.; Benoit, A.;
  Beretta, M.; Berg{\'e}, L.; Billard, J.; Borovlev, Y.A.; others.
\newblock {New Limit for Neutrinoless Double-Beta Decay of $^{100}$Mo from the
  CUPID-Mo Experiment}.
\newblock {\em Phys. Rev. Lett.} {\bf 2021}, {\em 126},~181802.

\bibitem[Augier \em{et~al.}(2022)Augier, Barabash, Bellini, Benato, Beretta,
  Berg\'e, Billard, Borovlev, Cardani, Casali, Cazes, et~al.]{Augier:2022}
Augier, C.; Barabash, A.; Bellini, F.; Benato, G.; Beretta, M.; Berg\'e, L.;
  Billard, J.; Borovlev, Y.; Cardani, L.; Casali, N.; Cazes, A.; others.
\newblock {Final results on the $0\nu\beta\beta$ decay half-life limit of
  $^{100}$Mo from the CUPID-Mo experiment}.
\newblock {\em Eur. Phys. J. C} {\bf 2022}, {\em 82},~1033.

\bibitem[Augier \em{et~al.}(2023)Augier, Barabash, Bellini, Benato, Beretta,
  Berg{\'e}, Billard, Borovlev, Cardani, Casali, et~al.]{Augier:2023a}
Augier, C.; Barabash, A.; Bellini, F.; Benato, G.; Beretta, M.; Berg{\'e}, L.;
  Billard, J.; Borovlev, Y.A.; Cardani, L.; Casali, N.; others.
\newblock {New measurement of double-$\beta$ decays of $^{100}$Mo to excited
  states of $^{100}$Ru with the CUPID-Mo experiment}.
\newblock {\em Phys. Rev. C} {\bf 2023}, {\em 107},~025503.

\bibitem[Armstrong \em{et~al.}(2019)Armstrong, Chang, Hafidi, Lisovenko,
  Novosad, Pearson, Polakovic, Wang, Yefremenko, Zhang,
  et~al.]{CUPIDInterestGroup:2019inu}
Armstrong, W.; Chang, C.; Hafidi, K.; Lisovenko, M.; Novosad, V.; Pearson, J.;
  Polakovic, T.; Wang, G.; Yefremenko, V.; Zhang, J.; others.
\newblock {CUPID pre-CDR}.
\newblock {\em arXiv} {\bf 2019},
  \href{http://xxx.lanl.gov/abs/1907.09376}{{\normalfont
  [arXiv:physics.ins-det/1907.09376]}}.

\bibitem[Armatol \em{et~al.}(2021{\natexlab{a}})Armatol, Armengaud, Armstrong,
  Augier, Avignone, Azzolini, Barabash, Bari, Barresi, Baudin,
  et~al.]{Armatol:2021a}
Armatol, A.; Armengaud, E.; Armstrong, W.; Augier, C.; Avignone, F.; Azzolini,
  O.; Barabash, A.; Bari, G.; Barresi, A.; Baudin, D.; others.
\newblock {Characterization of cubic Li$_2${}$^{100}$MoO$_4$ crystals for the
  CUPID experiment}.
\newblock {\em Eur. Phys. J. C} {\bf 2021}, {\em 81},~104.

\bibitem[Armatol \em{et~al.}(2021{\natexlab{b}})Armatol, Armengaud, Armstrong,
  Augier, Avignone~III, Azzolini, Bandac, Barabash, Bari, Barresi,
  et~al.]{Armatol:2021b}
Armatol, A.; Armengaud, E.; Armstrong, W.; Augier, C.; Avignone~III, F.;
  Azzolini, O.; Bandac, I.; Barabash, A.; Bari, G.; Barresi, A.; others.
\newblock {A CUPID Li$_2${}$^{100}$MoO$_4$ scintillating bolometer tested in
  the CROSS underground facility}.
\newblock {\em JINST} {\bf 2021}, {\em 16},~P02037.

\bibitem[Alfonso \em{et~al.}(2022)Alfonso, Armatol, Augier, Avignone~III,
  Azzolini, Balata, Barabash, Bari, Barresi, Baudin, et~al.]{Alfonso:2022}
Alfonso, K.; Armatol, A.; Augier, C.; Avignone~III, F.; Azzolini, O.; Balata,
  M.; Barabash, A.; Bari, G.; Barresi, A.; Baudin, D.; others.
\newblock {Optimization of the first CUPID detector module}.
\newblock {\em Eur. Phys. J. C} {\bf 2022}, {\em 82},~810.

\bibitem[Alfonso \em{et~al.}(2023)Alfonso, Armatol, Augier, Avignone~III,
  Azzolini, Balata, Bandac, Barabash, Bari, Barresi,
  et~al.]{CrossCupidTower:2023a}
Alfonso, K.; Armatol, A.; Augier, C.; Avignone~III, F.; Azzolini, O.; Balata,
  M.; Bandac, I.; Barabash, A.; Bari, G.; Barresi, A.; others.
\newblock {Twelve-crystal prototype of Li$_2$MoO$_4$ scintillating bolometers
  for CUPID and CROSS experiments}.
\newblock {\em Submitted to JINST} {\bf 2023},
  \href{http://xxx.lanl.gov/abs/2304.04611}{{\normalfont
  [arXiv:physics.ins-det/2304.04611]}}.

\bibitem[Bandac \em{et~al.}(2020)Bandac, Barabash, Berg{\'e}, Bri{\`e}re,
  Bourgeois, Carniti, Chapellier, de~Combarieu, Dafinei, Danevich,
  et~al.]{Bandac:2020}
Bandac, I.; Barabash, A.; Berg{\'e}, L.; Bri{\`e}re, M.; Bourgeois, C.;
  Carniti, P.; Chapellier, M.; de~Combarieu, M.; Dafinei, I.; Danevich, F.;
  others.
\newblock {The 0$\nu$2$\beta$-decay CROSS experiment: preliminary results and
  prospects}.
\newblock {\em JHEP} {\bf 2020}, {\em 01},~018.

\bibitem[Bandac \em{et~al.}(2021)Bandac, Barabash, Berg{\'e}, Bourgeois,
  Calvo-Mozota, Carniti, Chapellier, deCombarieu, Dafinei, Danevich,
  et~al.]{Bandac:2021}
Bandac, I.; Barabash, A.; Berg{\'e}, L.; Bourgeois, C.; Calvo-Mozota, J.;
  Carniti, P.; Chapellier, M.; deCombarieu, M.; Dafinei, I.; Danevich, F.;
  others.
\newblock {Phonon-mediated crystal detectors with metallic film coating capable
  of rejecting $\alpha$ and $\beta$ events induced by surface radioactivity}.
\newblock {\em Appl. Phys. Lett.} {\bf 2021}, {\em 118},~184105.

\bibitem[Alenkov \em{et~al.}(2015)Alenkov, Aryal, Beyer, Boiko, Boonin,
  Buzanov, Chanthima, Chernyak, Choi, Choi, et~al.]{Alenkov:2015}
Alenkov, V.; Aryal, P.; Beyer, J.; Boiko, R.; Boonin, K.; Buzanov, O.;
  Chanthima, N.; Chernyak, M.; Choi, J.; Choi, S.; others.
\newblock {Technical Design Report for the AMoRE $0\nu \beta \beta$ Decay
  Search Experiment}.
\newblock {\em arXiv} {\bf 2015},
  \href{http://xxx.lanl.gov/abs/1512.05957}{{\normalfont
  [arXiv:physics.ins-det/1512.05957]}}.

\bibitem[Kim \em{et~al.}(2020{\natexlab{a}})Kim, Jeon, Kim, Kim, Kim, Kim,
  Kwon, Lee, and So]{Kim:2020a}
Kim, H.; Jeon, J.; Kim, I.; Kim, S.; Kim, H.; Kim, Y.; Kwon, D.; Lee, M.; So,
  J.
\newblock {Compact phonon-scintillation detection system for rare event
  searches at low temperatures}.
\newblock {\em Nucl. Instrum. Methods Phys. Res. A} {\bf 2020}, {\em
  954},~162107.

\bibitem[Kim \em{et~al.}(2020{\natexlab{b}})Kim, Kim, Kim, Kim, Kim, Kim, Kwon,
  Jeon, Lee, Lee, et~al.]{Kim:2020b}
Kim, H.; Kim, H.; Kim, I.; Kim, S.R.; Kim, Y.; Kim, Y.H.; Kwon, D.; Jeon, J.A.;
  Lee, M.H.; Lee, M.; others.
\newblock {Li$_2$MoO$_4$ Phonon–Scintillation Detection Systems with MMC
  Readout}.
\newblock {\em J. Low Temp. Phys.} {\bf 2020}, {\em 199},~1082.

\bibitem[Kim \em{et~al.}(2022{\natexlab{a}})Kim, Kim, Sharma, Jeon, Kim, Kim,
  Kim, Kim, Kim, Lee, et~al.]{Kim:2022a}
Kim, W.; Kim, S.; Sharma, B.; Jeon, J.; Kim, H.; Kim, S.; Kim, S.; Kim, Y.;
  Kim, Y.; Lee, H.; others.
\newblock {Test Measurements of an MMC-Based 516-g Lithium Molybdate Crystal
  Detector for the AMoRE-II Experiment}.
\newblock {\em J. Low Temp. Phys.} {\bf 2022}, {\em 209},~299--307.

\bibitem[Kim \em{et~al.}(2022{\natexlab{b}})Kim, Ha, Jeon, Jeon, Jo, Kang,
  Kang, Kim, Kim, Kim, et~al.]{Kim:2022b}
Kim, H.; Ha, D.; Jeon, E.; Jeon, J.; Jo, H.; Kang, C.; Kang, W.; Kim, H.; Kim,
  S.; Kim, S.; others.
\newblock {Status and Performance of the AMoRE-I Experiment on Neutrinoless
  Double Beta Decay}.
\newblock {\em J. Low Temp. Phys.} {\bf 2022}, {\em 209},~962--970.

\bibitem[Bednyakov and \v{S}imkovic(2005)]{Bednyakov:2005}
Bednyakov, V.A.; \v{S}imkovic, F.
\newblock {Nuclear Spin Structure in Dark Matter Search: The Finite Momentum
  Transfer Limit}.
\newblock {\em Phys. Part. Nuclei} {\bf 2005}, {\em 37},~S106--S128.

\bibitem[Abdelhameed \em{et~al.}(2019)Abdelhameed, Angloher, Bauer, Bento,
  Bertoldo, Bucci, Canonica, D’Addabbo, Defay, Di~Lorenzo,
  et~al.]{Abdelhameed:2019a}
Abdelhameed, A.; Angloher, G.; Bauer, P.; Bento, A.; Bertoldo, E.; Bucci, C.;
  Canonica, L.; D’Addabbo, A.; Defay, X.; Di~Lorenzo, S.; others.
\newblock {First results on sub-GeV spin-dependent dark matter interactions
  with $^7$Li}.
\newblock {\em Eur. Phys. J. C} {\bf 2019}, {\em 79},~630.

\bibitem[Kr\v{c}mar \em{et~al.}(2001)Kr\v{c}mar, Kre\v{c}ak, Ljubi\v{c}i\v{c},
  Stip\v{c}evi\'{c}, and Bradley]{Krcmar:2001}
Kr\v{c}mar, M.; Kre\v{c}ak, Z.; Ljubi\v{c}i\v{c}, A.; Stip\v{c}evi\'{c}, M.;
  Bradley, D.A.
\newblock {Search for solar axions using ${}^{7}\mathrm{Li}$}.
\newblock {\em Phys. Rev. D} {\bf 2001}, {\em 64},~115016.

\bibitem[Belli \em{et~al.}(2008)Belli, Bernabei, Cerulli, Danevich, d'Angelo,
  Goriletsky, Grinyov, Incicchitti, Kobychev, Laubenstein, et~al.]{Belli:2008a}
Belli, P.; Bernabei, R.; Cerulli, R.; Danevich, F.; d'Angelo, A.; Goriletsky,
  V.; Grinyov, B.; Incicchitti, A.; Kobychev, V.; Laubenstein, M.; others.
\newblock {$^7$Li solar axions: Preliminary results and feasibility studies}.
\newblock {\em Nucl. Phys. A} {\bf 2008}, {\em 806},~388--397.

\bibitem[Barinova \em{et~al.}(2010)Barinova, Danevich, Degoda, Kirsanova,
  Kudovbenko, Pirro, and Tretyak]{Barinova:2010}
Barinova, O.; Danevich, F.; Degoda, V.Y.; Kirsanova, S.; Kudovbenko, V.; Pirro,
  S.; Tretyak, V.
\newblock First test of Li$_2$MoO$_4$ crystal as a cryogenic scintillating
  bolometer.
\newblock {\em Nucl. Instrum. Methods Phys. Res. A} {\bf 2010}, {\em
  613},~54--57.

\bibitem[Belli \em{et~al.}(2012)Belli, Bernabei, Cappella, Cerulli, Danevich,
  Incicchitti, Kobychev, Laubenstein, Polischuk, Tretyak, et~al.]{Belli:2012a}
Belli, P.; Bernabei, R.; Cappella, F.; Cerulli, R.; Danevich, F.; Incicchitti,
  A.; Kobychev, V.; Laubenstein, M.; Polischuk, O.; Tretyak, V.; others.
\newblock {Search for $^7$Li solar axions using resonant absorption in LiF
  crystal: Final results}.
\newblock {\em Phys. Lett. B} {\bf 2012}, {\em 711},~41--45.

\bibitem[Mart\'inez \em{et~al.}(2012)Mart\'inez, Coron, Ginestra, Gironnet,
  Gressier, Leblanc, de~Marcillac, Redon, Di~Stefano, Torres,
  et~al.]{Martinez:2012}
Mart\'inez, M.; Coron, N.; Ginestra, C.; Gironnet, J.; Gressier, V.; Leblanc,
  J.; de~Marcillac, P.; Redon, T.; Di~Stefano, P.; Torres, L.; others.
\newblock {Scintillating bolometers for fast neutron spectroscopy in rare
  events searches}.
\newblock {\em J. Phys.: Conf. Ser.} {\bf 2012}, {\em 375},~012025.

\bibitem[Coron \em{et~al.}(2016)Coron, Cuesta, Garc{\'\i}a, Ginestra, Gironnet,
  de~Marcillac, Mart{\'\i}nez, Ortigoza, de~Sol{\'o}rzano, Puimed{\'o}n,
  et~al.]{Coron:2016}
Coron, N.; Cuesta, C.; Garc{\'\i}a, E.; Ginestra, C.; Gironnet, J.;
  de~Marcillac, P.; Mart{\'\i}nez, M.; Ortigoza, Y.; de~Sol{\'o}rzano, A.O.;
  Puimed{\'o}n, J.; others.
\newblock {Neutron Spectrometry With Scintillating Bolometers of LiF and
  Sapphire}.
\newblock {\em IEEE Trans. Nucl. Sci.} {\bf 2016}, {\em 63},~1967--1975.

\bibitem[Grigorieva \em{et~al.}(2020)Grigorieva, Shlegel, Borovlev, Bekker,
  Barabash, Konovalov, Umatov, Borovkov, and Meshkov]{Grigorieva:2020}
Grigorieva, V.D.; Shlegel, V.; Borovlev, Y.; Bekker, T.; Barabash, A.;
  Konovalov, S.; Umatov, V.; Borovkov, V.; Meshkov, O.
\newblock {Li$_2${}$^{100\text{depl}}$MoO$_4$ crystals grown by
  low-thermal-gradient Czochralski technique}.
\newblock {\em J. Cryst. Growth} {\bf 2020}, {\em 552},~125913.

\bibitem[Berg\'e \em{et~al.}(2014)Berg\'e, Boiko, Chapellier, Chernyak, Coron,
  Danevich, Decourt, Degoda, Devoyon, Drillien, et~al.]{Berge:2014}
Berg\'e, L.; Boiko, R.; Chapellier, M.; Chernyak, D.; Coron, N.; Danevich, F.;
  Decourt, R.; Degoda, V.Y.; Devoyon, L.; Drillien, A.; others.
\newblock {Purification of molybdenum, growth and characterization of medium
  volume ZnMoO$_4$ crystals for the LUMINEU program}.
\newblock {\em JINST} {\bf 2014}, {\em 9},~P06004.

\bibitem[Berg{\'e} \em{et~al.}(2018)Berg{\'e}, Chapellier, De~Combarieu,
  Dumoulin, Giuliani, Gros, De~Marcillac, Marnieros, Nones, Novati,
  et~al.]{Berge:2018}
Berg{\'e}, L.; Chapellier, M.; De~Combarieu, M.; Dumoulin, L.; Giuliani, A.;
  Gros, M.; De~Marcillac, P.; Marnieros, S.; Nones, C.; Novati, V.; others.
\newblock {Complete event-by-event $\alpha$/$\gamma$($\beta$) separation in a
  full-size TeO$_2$ CUORE bolometer by Neganov-Luke-magnified light detection}.
\newblock {\em Phys. Rev. C} {\bf 2018}, {\em 97},~032501.

\bibitem[Haller(1994)]{Haller:1994}
Haller, E.E.
\newblock {Advanced far-infrared detectors}.
\newblock {\em Infrared Phys. Techn.} {\bf 1994}, {\em 35},~127.

\bibitem[Andreotti \em{et~al.}(2012)Andreotti, Brofferio, Foggetta, Giuliani,
  Margesin, Nones, Pedretti, Rusconi, Salvioni, and Tenconi]{Andreotti:2012}
Andreotti, E.; Brofferio, C.; Foggetta, L.; Giuliani, A.; Margesin, B.; Nones,
  C.; Pedretti, M.; Rusconi, C.; Salvioni, C.; Tenconi, M.
\newblock {Production, characterization and selection of the heating elements
  for the response stabilization of the CUORE bolometers}.
\newblock {\em Nucl. Instrum. Meth. A} {\bf 2012}, {\em 664},~161.

\bibitem[Alessandrello \em{et~al.}(1998)Alessandrello, Brofferio, Bucci,
  Cremonesi, Giuliani, Margesin, Nucciotti, Pavan, Pessina, Previtali,
  et~al.]{Alessandrello:1998}
Alessandrello, A.; Brofferio, C.; Bucci, C.; Cremonesi, O.; Giuliani, A.;
  Margesin, B.; Nucciotti, A.; Pavan, M.; Pessina, G.; Previtali, E.; others.
\newblock {Methods for response stabilization in bolometers for rare decays}.
\newblock {\em J. Cryst. Growth} {\bf 1998}, {\em 412},~454.

\bibitem[Azzolini \em{et~al.}(2018)Azzolini, Barrera, Beeman, Bellini, Beretta,
  Biassoni, Brofferio, Bucci, Canonica, Capelli, et~al.]{Azzolini:2018tum}
Azzolini, O.; Barrera, M.; Beeman, J.; Bellini, F.; Beretta, M.; Biassoni, M.;
  Brofferio, C.; Bucci, C.; Canonica, L.; Capelli, S.; others.
\newblock {CUPID-0: the first array of enriched scintillating bolometers for
  $0\nu\beta\beta$ decay investigations}.
\newblock {\em Eur. Phys. J. C} {\bf 2018}, {\em 78},~428.

\bibitem[Armatol \em{et~al.}(2021)Armatol, Armengaud, Armstrong, Augier,
  Avignone~III, Azzolini, Barabash, Bari, Barresi, Baudin,
  et~al.]{Armatol:2021}
Armatol, A.; Armengaud, E.; Armstrong, W.; Augier, C.; Avignone~III, F.;
  Azzolini, O.; Barabash, A.; Bari, G.; Barresi, A.; Baudin, D.; others.
\newblock {Novel technique for the study of pileup events in cryogenic
  bolometers}.
\newblock {\em Phys. Rev. C} {\bf 2021}, {\em 104},~015501.

\bibitem[Umi()]{Umicore}
Umicore Germanium substrates.
\newblock
  \url{https://eom.umicore.com/en/germanium-solutions/products/germanium-substrates/}.
\newblock Accessed: 2023-04-25.

\bibitem[Mancuso \em{et~al.}(2014)Mancuso, Beeman, Giuliani, Dumoulin,
  Olivieri, Pessina, Plantevin, Rusconi, and Tenconi]{Mancuso:2014}
Mancuso, M.; Beeman, J.; Giuliani, A.; Dumoulin, L.; Olivieri, E.; Pessina, G.;
  Plantevin, O.; Rusconi, C.; Tenconi, M.
\newblock {An experimental study of antireflective coatings in Ge light
  detectors for scintillating bolometers}.
\newblock {\em EPJ Web Conf.} {\bf 2014}, {\em 65},~04003.

\bibitem[Olivieri and {CROSS collaboration}(2020)]{Olivieri:2020}
Olivieri, E.; {CROSS collaboration}.
\newblock {The new CROSS Cryogenic Underground (C2U) facility: an overview}.
\newblock  {XXIX International (online) Conference on Neutrino Physics and
  Astrophysics (Neutrino 2020), June 22 -- July 02, 2020},  2020.
\newblock {Poster presented at the XXIX International (online) Conference on
  Neutrino Physics and Astrophysics (Neutrino 2020), June 22 -- July 02, 2020}.

\bibitem[Trzaska \em{et~al.}(2019)Trzaska, Slupecki, Bandac, Bayo, Bettini,
  Bezrukov, Enqvist, Fazliakhmetov, Ianni, Inzhechik, et~al.]{Trzaska:2019}
Trzaska, W.H.; Slupecki, M.; Bandac, I.; Bayo, A.; Bettini, A.; Bezrukov, L.;
  Enqvist, T.; Fazliakhmetov, A.; Ianni, A.; Inzhechik, L.; others.
\newblock {Cosmic-ray muon flux at Canfranc Underground Laboratory}.
\newblock {\em Eur. Phys. J. C} {\bf 2019}, {\em 79},~721.

\bibitem[UQT()]{UQT}
The Ultra-Quiet Technology.
\newblock \url{https://cryoconcept.com/product/the-ultra-quiet-technology/}.
\newblock Accessed: 2023-04-25.

\bibitem[Olivieri \em{et~al.}(2017)Olivieri, Billard, De~Jesus, Juillard, and
  Leder]{Olivieri:2017}
Olivieri, E.; Billard, J.; De~Jesus, M.; Juillard, A.; Leder, A.
\newblock {Vibrations on pulse tube based Dry Dilution Refrigerators for low
  noise measurements}.
\newblock {\em Nucl. Instrum. Meth. A} {\bf 2017}, {\em 858},~73.

\bibitem[Khalife(2021)]{Khalife:2021}
Khalife, H.
\newblock {CROSS and CUPID-Mo : future strategies and new results in bolometric
  search for $0\nu\beta\beta$}.
\newblock PhD thesis, Universit\'e Paris-Saclay, Orsay, France,  2021.

\bibitem[Vel{\'a}zquez \em{et~al.}(2017)Vel{\'a}zquez, Veber, Moutatouia,
  De~Marcillac, Giuliani, Loaiza, Denux, Decourt, El~Hafid, Laubenstein,
  et~al.]{Velazquez:2017}
Vel{\'a}zquez, M.; Veber, P.; Moutatouia, M.; De~Marcillac, P.; Giuliani, A.;
  Loaiza, P.; Denux, D.; Decourt, R.; El~Hafid, H.; Laubenstein, M.; others.
\newblock {Exploratory growth in the Li$_2$MoO$_4$-MoO$_3$ system for the next
  crystal generation of heat-scintillation cryogenic bolometers}.
\newblock {\em Solid State Sci.} {\bf 2017}, {\em 65},~41.

\bibitem[Stelian \em{et~al.}(2020)Stelian, Velazquez, Veber, Ahmine, Duffar,
  de~Marcillac, Giuliani, Poda, Marnieros, Nones, et~al.]{Stelian:2020}
Stelian, C.; Velazquez, M.; Veber, P.; Ahmine, A.; Duffar, T.; de~Marcillac,
  P.; Giuliani, A.; Poda, D.; Marnieros, S.; Nones, C.; others.
\newblock {Experimental and numerical investigations of the Czochralski growth
  of Li$_2$MoO$_4$ crystals for heat-scintillation cryogenic bolometers}.
\newblock {\em J. Cryst. Growth} {\bf 2020}, {\em 531},~125385.

\bibitem[Zolotarova and {CROSS collaboration}(2020)]{Zolotarova:2020}
Zolotarova, A.; {CROSS collaboration}.
\newblock The {CROSS} experiment: search for 0$\nu$2$\beta$ decay with surface
  sensitive bolometers.
\newblock {\em J. Phys.: Conf. Ser.} {\bf 2020}, {\em 1468},~012147.

\bibitem[Helis \em{et~al.}(2020)Helis, Bandac, Barabash, Billard, Chapellier,
  de~Combarieu, Danevich, Dumoulin, Gascon, Giuliani, Kasperovych,
  et~al.]{Helis:2020}
Helis, D.L.; Bandac, I.; Barabash, A.; Billard, J.; Chapellier, M.;
  de~Combarieu, M.; Danevich, F.; Dumoulin, L.; Gascon, J.; Giuliani, A.;
  Kasperovych, D.; others.
\newblock {Neutrinoless double-beta decay searches with enriched
  $^{116}$CdWO$_4$ scintillating bolometers}.
\newblock {\em J. Low Temp. Phys.} {\bf 2020}, {\em 199},~467.

\bibitem[Helis(2021)]{Helis:2021}
Helis, D.
\newblock {Searching for neutrinoless double-beta decay with scintillating
  bolometers}.
\newblock PhD thesis, Universit\'e Paris-Saclay, Orsay, France,  2021.

\bibitem[Arnaboldi \em{et~al.}(2002)Arnaboldi, Bucci, Campbell, Capelli,
  Nucciotti, Pavan, Pessina, Pirro, Previtali, Rosenfeld,
  et~al.]{Arnaboldi:2002}
Arnaboldi, C.; Bucci, C.; Campbell, J.; Capelli, S.; Nucciotti, A.; Pavan, M.;
  Pessina, G.; Pirro, S.; Previtali, E.; Rosenfeld, C.; others.
\newblock The programmable front-end system for {CUORICINO}, an array of
  large-mass bolometers.
\newblock {\em IEEE Trans. Nucl. Sci.} {\bf 2002}, {\em 49},~2440--2447.

\bibitem[Carniti \em{et~al.}(2020)Carniti, Gotti, and Pessina]{Carniti:2020}
Carniti, P.; Gotti, C.; Pessina, G.
\newblock {High-Resolution Digitization System for the CROSS Experiment}.
\newblock {\em J. Low Temp. Phys.} {\bf 2020}, {\em 199},~833.

\bibitem[Carniti \em{et~al.}(2023)Carniti, Gotti, and Pessina]{Carniti:2023}
Carniti, P.; Gotti, C.; Pessina, G.
\newblock {High resolution filtering and digitization system for cryogenic
  bolometric detectors}.
\newblock {\em Nucl. Instrum. Meth. A} {\bf 2023}, {\em 1045},~167658.

\bibitem[Novati \em{et~al.}(2019)Novati, Berg{\'e}, Dumoulin, Giuliani,
  Mancuso, de~Marcillac, Marnieros, Olivieri, Poda, Tenconi,
  et~al.]{Novati:2019}
Novati, V.; Berg{\'e}, L.; Dumoulin, L.; Giuliani, A.; Mancuso, M.;
  de~Marcillac, P.; Marnieros, S.; Olivieri, E.; Poda, D.; Tenconi, M.; others.
\newblock {Charge-to-heat transducers exploiting the Neganov-Trofimov-Luke
  effect for light detection in rare-event searches}.
\newblock {\em Nucl. Instrum. Methods Phys. Res. A} {\bf 2019}, {\em
  940},~320--327.

\bibitem[Mancuso(2016)]{Mancuso:2016}
Mancuso, M.
\newblock Development and optimization of scintillating bolometers and
  innovative light detectors for the search for neutrinoless double beta decay.
\newblock PhD thesis, Universit\'e Paris-Sud, Orsay, France,  2016.

\bibitem[Gatti and Manfredi(1986)]{Gatti:1986}
Gatti, E.; Manfredi, P.
\newblock Processing the signals from solid-state detectors in
  elementary-particle physics.
\newblock {\em Riv. Nuovo Cim.} {\bf 1986}, {\em 9},~1.

\bibitem[Beeman \em{et~al.}(2013)Beeman, Bellini, Casali, Cardani, Dafinei,
  Di~Domizio, Ferroni, Gironi, Nagorny, Orio, et~al.]{Beeman:2013b}
Beeman, J.; Bellini, F.; Casali, N.; Cardani, L.; Dafinei, I.; Di~Domizio, S.;
  Ferroni, F.; Gironi, L.; Nagorny, S.; Orio, F.; others.
\newblock {Characterization of bolometric Light Detectors for rare event
  searches}.
\newblock {\em JINST} {\bf 2013}, {\em 8},~P07021.

\bibitem[Pirro \em{et~al.}(2006)Pirro, Beeman, Capelli, Pavan, Previtali, and
  Gorla]{Pirro:2005ar}
Pirro, S.; Beeman, J.; Capelli, S.; Pavan, M.; Previtali, E.; Gorla, P.
\newblock {Scintillating double beta decay bolometers}.
\newblock {\em Phys. Atom. Nucl.} {\bf 2006}, {\em 69},~2109.

\bibitem[Piperno \em{et~al.}(2011)Piperno, Pirro, and Vignati]{Piperno:2011}
Piperno, G.; Pirro, S.; Vignati, M.
\newblock {Optimizing the energy threshold of light detectors coupled to
  luminescent bolometers}.
\newblock {\em JINST} {\bf 2011}, {\em 6},~P10005.

\bibitem[Poda and {CUPID-Mo collaboration}(2020)]{Poda:2020}
Poda, D.V.; {CUPID-Mo collaboration}.
\newblock {Performance of the CUPID-Mo double-beta decay bolometric
  experiment}.
\newblock  {XXIX International (online) Conference on Neutrino Physics and
  Astrophysics (Neutrino 2020), June 22 -- July 02, 2020},  2020.
\newblock {Poster presented at the XXIX International (online) Conference on
  Neutrino Physics and Astrophysics (Neutrino 2020), June 22 -- July 02, 2020}.

\bibitem[Armengaud \em{et~al.}(2015)Armengaud et~al.]{Armengaud:2015}
Armengaud, E.; others.
\newblock {Development and underground test of radiopure ZnMoO$_4$
  scintillating bolometers for the LUMINEU $0\nu2\beta$ project}.
\newblock {\em JINST} {\bf 2015}, {\em 10},~P05007.

\bibitem[Danevich(2017)]{Danevich:2017}
Danevich, F.A.
\newblock {Radiopure tungstate and molybdate crystal scintillators for double
  beta decay experiments}.
\newblock {\em Int. J. Mod. Phys. A} {\bf 2017}, {\em 32},~1743008.

\bibitem[Danevich and Tretyak(2018)]{Danevich:2018}
Danevich, F.A.; Tretyak, V.I.
\newblock {Radioactive contamination of scintillators}.
\newblock {\em Int. J. Mod. Phys. A} {\bf 2018}, {\em 33},~1843007.

\bibitem[Feldman and Cousins(1998)]{Feldman:1998}
Feldman, G.J.; Cousins, R.D.
\newblock {Unified approach to the classical statistical analysis of small
  signals}.
\newblock {\em Phys. Rev. D} {\bf 1998}, {\em 3873},~57.

\bibitem[Barabash(2020)]{Barabash:2020}
Barabash, A.S.
\newblock {Precise Half-Life Values for Two-Neutrino Double-$\beta$ Decay: 2020
  Review}.
\newblock {\em Universe} {\bf 2020}, {\em 6},~159.

\bibitem[Augier \em{et~al.}(2023)Augier, Barabash, Bellini, Benato, Beretta,
  Berg\'e, Billard, Borovlev, Cardani, Casali, Cazes,
  et~al.]{CUPIDMoBkgModel:2023}
Augier, C.; Barabash, A.; Bellini, F.; Benato, G.; Beretta, M.; Berg\'e, L.;
  Billard, J.; Borovlev, Y.; Cardani, L.; Casali, N.; Cazes, A.; others.
\newblock {The background model of the CUPID-Mo $0\nu\beta\beta$ experiment}.
\newblock {\em Eur. Phys. J. C (to be submitted)} {\bf 2023}.

\bibitem[Azzolini \em{et~al.}(2019)Azzolini, Beeman, Bellini, Beretta,
  Biassoni, Brofferio, Bucci, Capelli, Cardani, Carniti,
  et~al.]{Azzolini:2019nmi}
Azzolini, O.; Beeman, J.; Bellini, F.; Beretta, M.; Biassoni, M.; Brofferio,
  C.; Bucci, C.; Capelli, S.; Cardani, L.; Carniti, P.; others.
\newblock {Background model of the CUPID-0 experiment}.
\newblock {\em Eur. Phys. J. C} {\bf 2019}, {\em 79},~583.

\bibitem[Azzolini \em{et~al.}(2022)Azzolini, Beeman, Bellini, Beretta,
  Biassoni, Brofferio, Bucci, Capelli, Caracciolo, Cardani,
  et~al.]{Azzolini:2022}
Azzolini, O.; Beeman, J.; Bellini, F.; Beretta, M.; Biassoni, M.; Brofferio,
  C.; Bucci, C.; Capelli, S.; Caracciolo, V.; Cardani, L.; others.
\newblock {Final Result on the Neutrinoless Double Beta Decay of
  $^{82}\mathrm{Se}$ with CUPID-0}.
\newblock {\em Phys. Rev. Lett.} {\bf 2022}, {\em 129},~111801.

\bibitem[Adams \em{et~al.}(2021)Adams, Alduino, Alfonso, Avignone~III,
  Azzolini, Bari, Bellini, Benato, Biassoni, Branca, et~al.]{Adams:2021}
Adams, D.; Alduino, C.; Alfonso, K.; Avignone~III, F.; Azzolini, O.; Bari, G.;
  Bellini, F.; Benato, G.; Biassoni, M.; Branca, A.; others.
\newblock {Measurement of the $2\nu\beta\beta$ Decay Half-Life of $^{130}$Te
  with CUORE}.
\newblock {\em Phys. Rev. Lett.} {\bf 2021}, {\em 126},~171801.

\bibitem[Adams \em{et~al.}(2022)Adams, Alduino, Alfonso, Avignone~III,
  Azzolini, Bari, Bellini, Benato, Beretta, Biassoni, et~al.]{Adams:2022}
Adams, D.; Alduino, C.; Alfonso, K.; Avignone~III, F.; Azzolini, O.; Bari, G.;
  Bellini, F.; Benato, G.; Beretta, M.; Biassoni, M.; others.
\newblock {Search for Majorana neutrinos exploiting millikelvin cryogenics with
  CUORE}.
\newblock {\em Nature} {\bf 2022}, {\em 604},~53.

\bibitem[{der~}Mateosian and Goldhaber(1966)]{Mateosian:1966}
{der~}Mateosian, E.; Goldhaber, M.
\newblock {Limits for Lepton-Conserving and Lepton-Nonconserving Double Beta
  Decay in Ca$^{48}$}.
\newblock {\em Phys. Rev.} {\bf 1966}, {\em 146},~810--815.

\bibitem[Gavrilyuk \em{et~al.}(2013)Gavrilyuk, Gangapshev, Kazalov, Kuzminov,
  Panasenko, and Ratkevich]{Gavrilyuk:2013}
Gavrilyuk, Y.M.; Gangapshev, A.M.; Kazalov, V.V.; Kuzminov, V.V.; Panasenko,
  S.I.; Ratkevich, S.S.
\newblock {Indications of $2\nu2K$ capture in $^{78}$Kr}.
\newblock {\em Phys. Rev. C} {\bf 2013}, {\em 87},~035501.

\bibitem[Novella \em{et~al.}(2022)Novella, Sorel, Us{\'o}n, Adams, Almaz{\'a}n,
  {\'A}lvarez, Aparicio, Aranburu, Arazi, Arnquist, et~al.]{Novella:2022}
Novella, P.; Sorel, M.; Us{\'o}n, A.; Adams, C.; Almaz{\'a}n, H.; {\'A}lvarez,
  V.; Aparicio, B.; Aranburu, A.; Arazi, L.; Arnquist, I.; others.
\newblock {Measurement of the $^{136}$Xe two-neutrino double-$\beta$-decay
  half-life via direct background subtraction in NEXT}.
\newblock {\em Phys. Rev. C} {\bf 2022}, {\em 105},~055501.

\bibitem[Giuliani and Poves(2012)]{Giuliani:2012a}
Giuliani, A.; Poves, A.
\newblock {Neutrinoless Double-Beta Decay}.
\newblock {\em Adv. High Energy Phys.} {\bf 2012}, {\em 2012},~857016.

\end{thebibliography}
\end{document}